\documentclass[twocolumn,showpacs,preprintnumbers,amsmath,amssymb]{revtex4}
\usepackage{graphicx}
\usepackage{dcolumn}
\usepackage{bm}
\usepackage{color}
\usepackage{amsmath,amsfonts,amssymb}
\usepackage{float}
\usepackage{verbatim}
\usepackage{bm}
\usepackage{mathtools}
\usepackage{anyfontsize}
\usepackage{subfigure}

\begin{document}

\title{Densification and Structural Transitions in Networks that Grow by Node
  Copying}

\author{U. Bhat}
\affiliation{Department of Physics, Boston University, Boston, Massachusetts 02215, USA}
\affiliation{Santa Fe Institute, 1399 Hyde Park Road, Santa Fe, NM, 87501}
\author{P. L. Krapivsky}
\affiliation{Department of Physics, Boston University, Boston, Massachusetts 02215, USA}
\author{R. Lambiotte}
\affiliation{naXys, Namur Center for Complex Systems, University of Namur,
rempart de la Vierge 8, B 5000 Namur, Belgium}
\author{S. Redner}
\affiliation{Santa Fe Institute, 1399 Hyde Park Road, Santa Fe, NM, 87501}

\begin{abstract}
  We introduce a growing network model---the copying model---in which
  a new node attaches to a randomly selected target node and, in addition,
  independently to each of the neighbors of the target with copying
  probability $p$.  When $p<\frac{1}{2}$, this algorithm generates sparse networks,
  in which the average node degree is finite.  A power-law degree
  distribution also arises, with a non-universal exponent whose value is
  determined by a transcendental equation in $p$.  In the sparse regime, the network is
  ``normal'', e.g., the relative fluctuations in the number of links are asymptotically
  negligible.  For $p\geq \frac{1}{2}$, the emergent networks are dense 
  (the average degree increases with the number of nodes $N$) and they exhibit 
   intriguing structural behaviors.  In particular, the
  $N$-dependence of the number of $m$-cliques (complete subgraphs of $m$
  nodes) undergoes $m-1$ transitions from normal to progressively more
  anomalous behavior at a $m$-dependent critical values of $p$.  Different
  realizations of the network, which start from the same initial state,
  exhibit macroscopic fluctuations in the thermodynamic limit---absence of
  self averaging.  When linking to second neighbors of the target node can
  occur, the number of links asymptotically grows as $N^2$ as $N\to\infty$,
  so that the network is effectively complete as $N\to \infty$.

\end {abstract}

\maketitle

\section{Introduction and Model}

A wide variety of complex networks grow by copying mechanisms.  As examples,
copying and redirection are key ingredients in the growth of the world-wide
web, citation networks and other information networks
\cite{GNC,sergi,kleinberg,evans}.  In social networks, copying corresponds to
triadic closure, that is, the formation of new social ties between two
friends of a given individual.  This mechanism appears to be important in
driving social network dynamics~\cite{granovetter,toivonen}.  Copying also
occurs in Nature.  For example, the process of gene duplication, which is
essentially the copying mechanism, plays a crucial role in
evolution~\cite{japan,bio}.  Various models for
protein interaction networks are also based on duplication and
divergence~\cite{pastor,chung,vesp,korea,wag,nik,protein,Makse,sym,dupl+biol,dupl+math}.

From a modeling viewpoint, the copying mechanism has the advantage of being
local~\cite{vaz,GNR,saram}, as the creation of new links only depends on the
nearest neighborhood of each node, in contrast to global rules, such as
preferential attachment.  Despite the simplicity of the copying rule, there
has not yet been a a rigorous mathematical analysis of networks that are
generated by this mechanism.

\begin{figure}[ht]
\begin{center}
\includegraphics[width=0.4\textwidth]{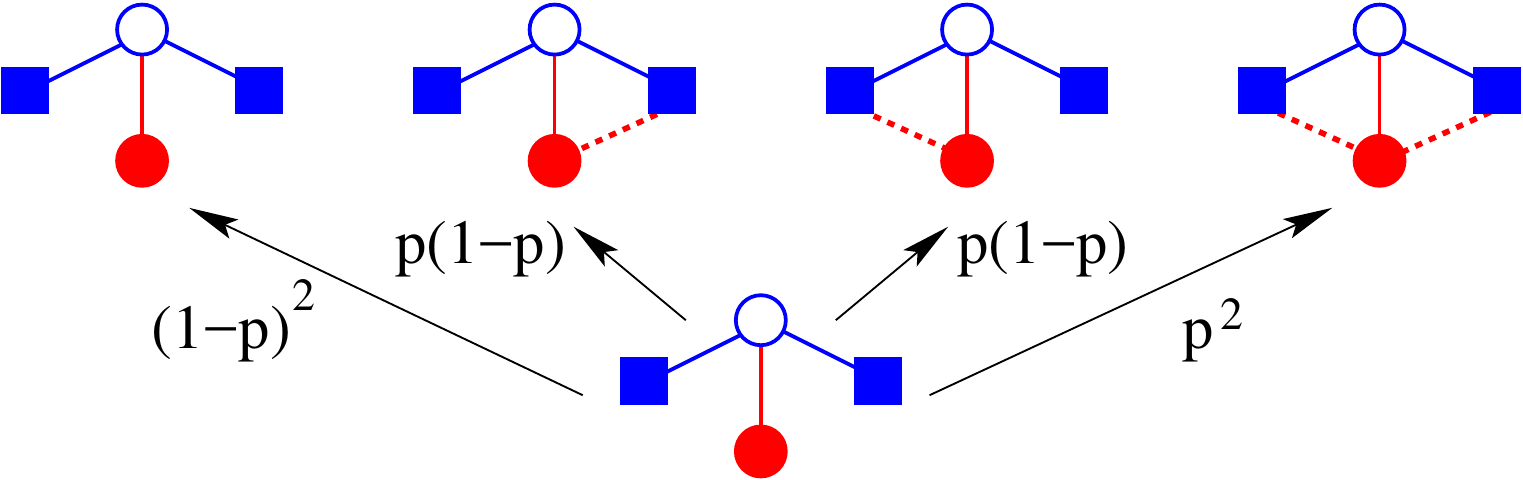}
\caption{\small The copying model.  A new node (filled circle) attaches to a
  random target (open circle) and independently to each of the friends of the
  target (squares) with probability $p$. }
\label{cartoon}
\end{center}
\end{figure}

\begin{figure*}[ht]
\hskip 0.4in\includegraphics[width=0.15\textwidth]{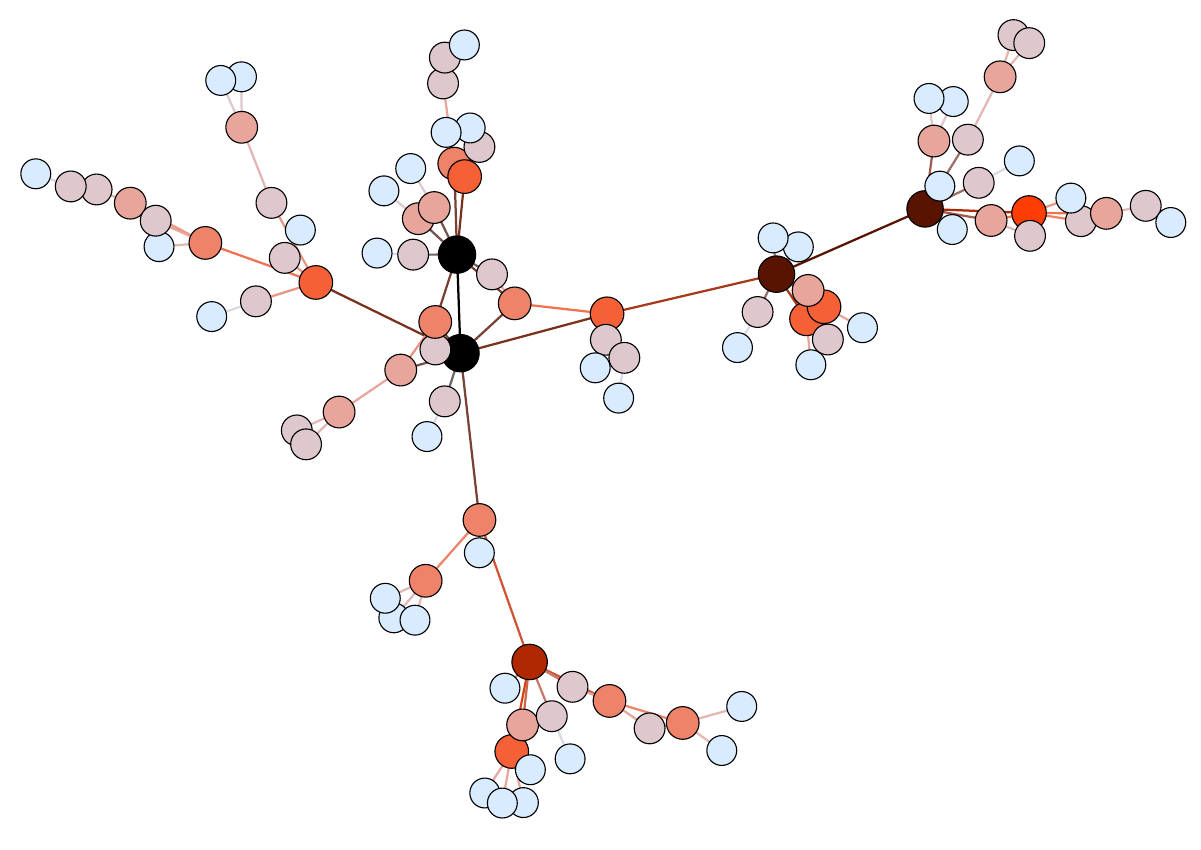}
\hskip 0.15in\includegraphics[width=0.15\textwidth]{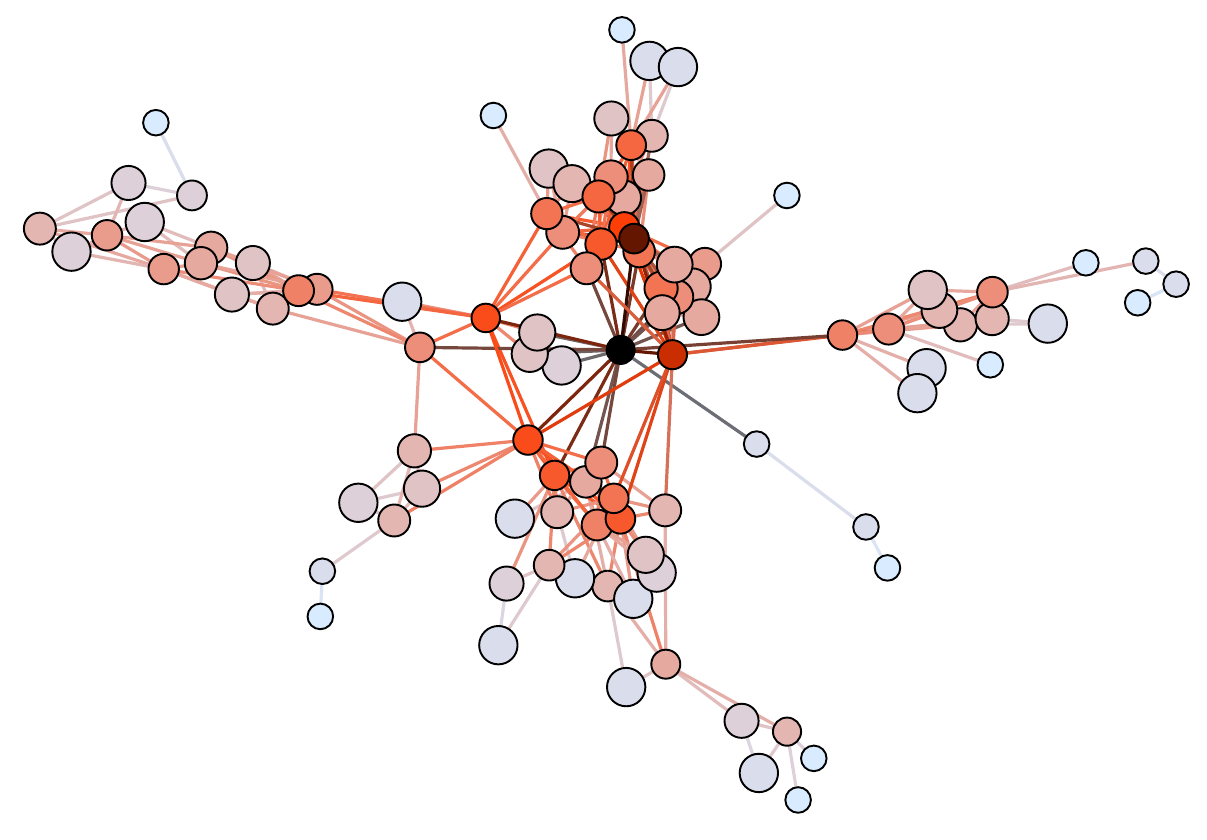}
\hskip 0.3in\includegraphics[width=0.12\textwidth]{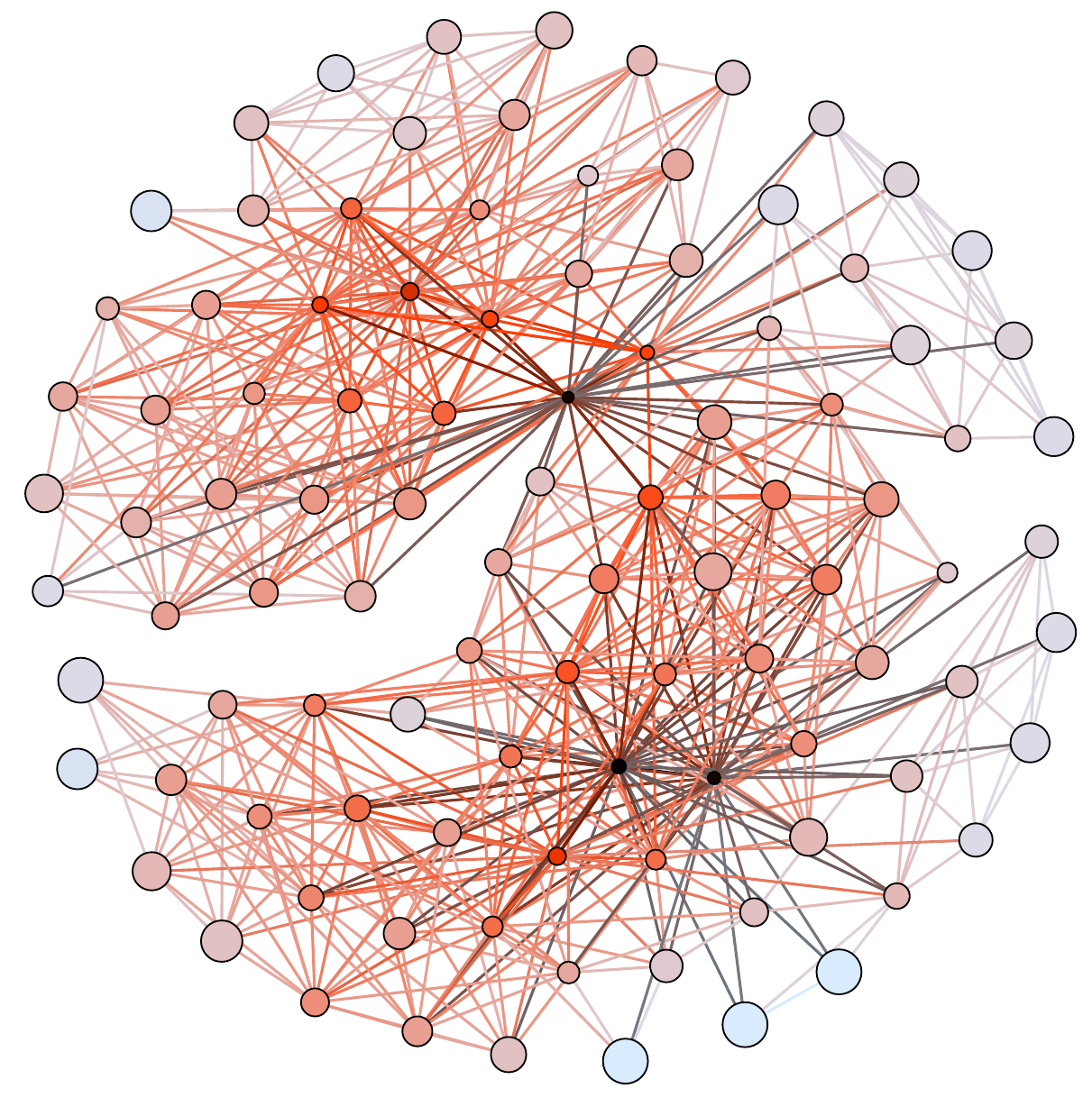}
\hskip 0.4in\includegraphics[width=0.12\textwidth]{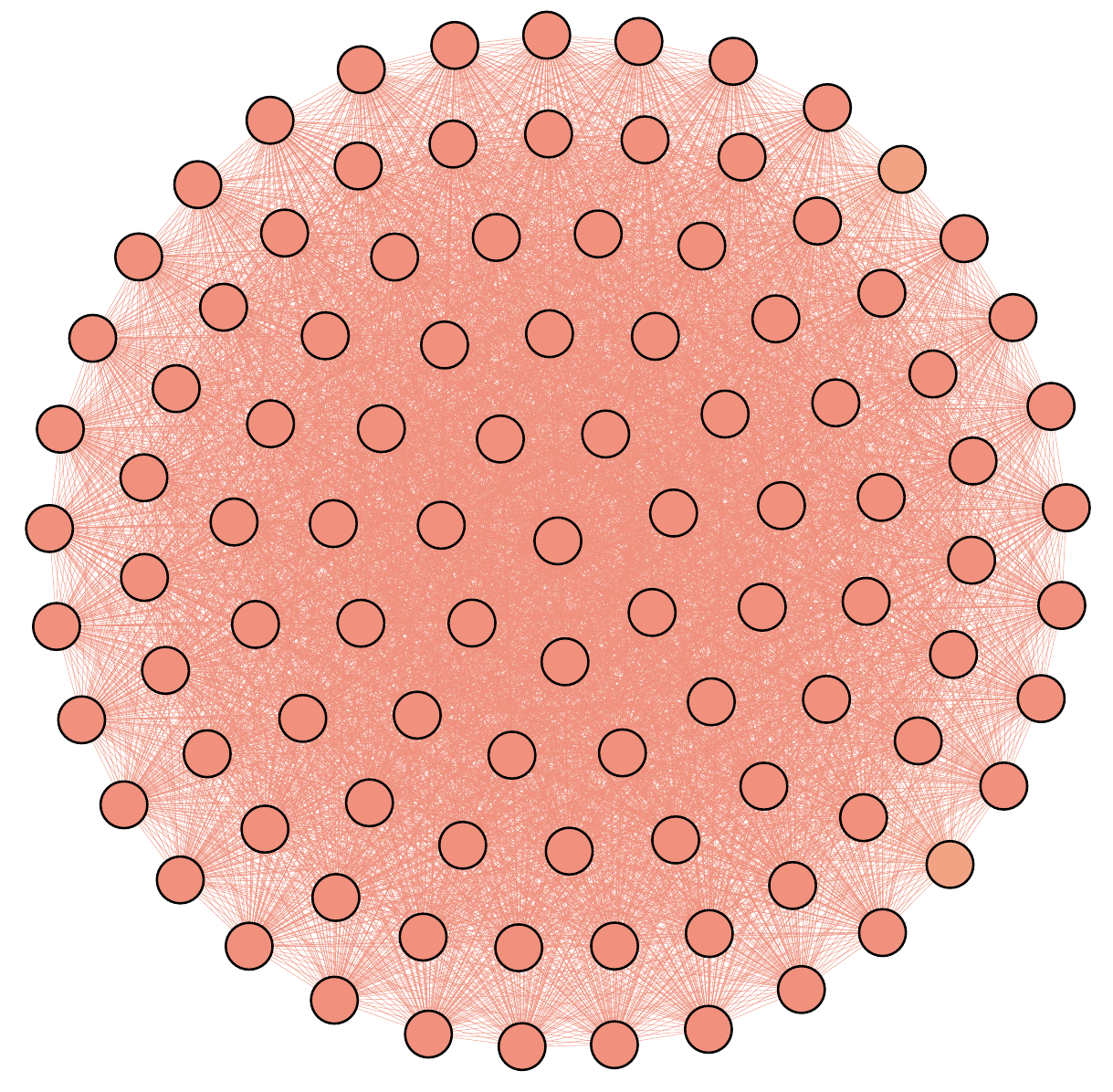}
\hbox{\hskip 1.1in\includegraphics[width=0.8\textwidth]{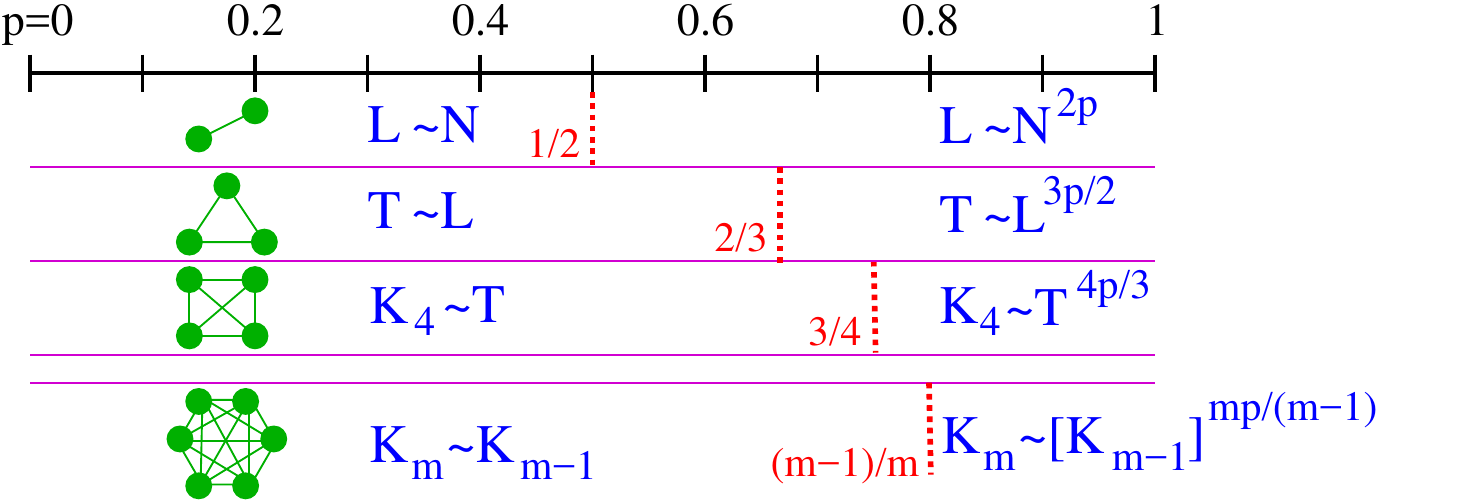}}
\caption{Realizations of the copying model for $p=0.1$, $0.4$, 0.7, and 1 for
  $N=100$, and a summary of the dense regimes.  For simplicity, only the last
  of the structural transitions is shown (see Sec.~\ref{sec:cliques}. }
\label{vis}
\end{figure*}

Here we investigate networks that grow by an elementary implementation of the
copying mechanism, which depends on only a single parameter---the copying
probability $p$ (Fig.~\ref{cartoon}).  In our copying model, a network
grows by adding nodes sequentially.  Each new node connects to a randomly
chosen target node and, in addition, independently to each of the neighbors
of the target with probability $p$.  The simplicity of this growth mechanism
allows us to develop an analytical description of many of the rich network
properties that emerge.

Perhaps the most crucial structural change is the transition from sparse
networks for $p<\frac{1}{2}$, where the number of links $L_N$ in a network of
$N$ nodes grows linearly with $N$, to dense networks, where $L_N$ grows
super-linearly with $N$.  In the sparse regime, the network is ``normal'' in
the sense that a typical realization of the network is representative of the
average behavior.  In contrast, in the dense regime, $p>\frac{1}{2}$, network
growth is not self-averaging; namely sample-to-sample fluctuations do not
vanish even when the number of nodes $N$ is very large.  In addition, the
copying model undergoes infinitely many transitions at
$p=\tfrac{2}{3}, \tfrac{3}{4}, \tfrac{4}{5},\ldots$ where sudden changes
arise in the growth laws of the number of triangles and progressively
higher-order cliques---complete subgraphs of $m$ nodes.  Moreover, for
intermediate values of $p$, the resulting networks appear to be highly
clustered (Fig.~\ref{vis}).

This article is organized as follows.  In the next Sec.~\ref{sec:links} we
quantify the simplest global network characteristic, the number of links
$L_N$.  Specifically, we show that the $N$-dependence of the average number
of links has a transition point at $p=\frac{1}{2}$, while the variance of
$L_N$ has transition points at $p=\frac{1}{4}$ and at $p=\frac{1}{2}$.  We
then analyze the degree distribution in Sec.~\ref{degree}, and show that it
has a power-law tail with a non-universal exponent in the sparse regime.  In
the dense regime, nearly all features of the degree distribution are
anomalous.  In Sec.~\ref{sec:cliques} we determine the growth laws for the
average number of triangles and higher-order $m$-cliques.  Cliques undergo a
rich sequence of structural transitions as $p$ increases.  In
Sec.~\ref{sec:clustering}, we analyze the clustering properties of the
network as a function of the copying probability $p$ and argue that maximal
cluster occurs at an intermediate value of $p$. In Sec.~\ref{PLN}, we examine
the probability distributions for the number of links $L_N$ and triangles
$T_N$.  In Sec.~\ref{sec:2nd-neighbor} we briefly discuss what happens if, in
addition to connecting to the neighbors of the target node, connections to
second neighbors are also allowed.  Finally in Sec.~\ref{concl}, we conclude
and discuss some possible open questions.

\section{Number of Links}
\label{sec:links}

A basic global characteristic of a network of $N$ nodes is the number of
links $L_N$.  In many models, the dependence of $L_N$ on $N$ is trivial.  For
example, if each new node links to $m$ pre-exiting nodes, then
$L_N = m(N-1)$. Hereinafter we assume that the network starts with a single node, so that
$L_1=0$. In the copying model, however, $L_N$ is a random variable taking
different values in different realizations. The exceptions are the extreme cases
of $p=0$ and $p=1$ where the number of links is deterministic. 
In the former case the copying mechanism produces a tree 
(more precisely, a random recursive tree), so $L_N = N-1$. When $p=1$, the copying model leads to 
the complete graph which has $L_N=N(N-1)/2$ links.

\subsection{The average $L(N)$}
\label{average}

The simplest characterization of the random quantity $L_N$ is the
\emph{average} number of links $L(N) \equiv \langle L_N \rangle$.  When a new
node is added, the average number of links increases by
$1+p\langle k\rangle$, where $\langle k\rangle=2L(N)/N$ is the average node
degree.  The factor 1 accounts for direct linking and the factor
$p\langle k\rangle$ accounts for copying events. Indeed, for a target node of
degree $k$, $pk$ additional links are created on average by copying
(Fig.~\ref{cartoon}).  Thus the average number of links grows as
\begin{align}
\label{LN-exact}
L(N\!+\!1)
&= \Big(1+\frac{2p}{N}\Big)L(N)+1\,.
\end{align}
Since we assume that the network starts with a single node, the initial condition is $L(1)=0$.

The solution to the homogeneous version of recursion \eqref{LN-exact} is
elementary.  Using this solution as a integrating factor, we solve the
inhomogeneous equation (see Appendix~\ref{app:LN}), from which the asymptotic
behavior is
\begin{subequations}
\label{L-all}
\begin{equation}
\label{LN-sol}
L(N)= 
\begin{cases}
{\displaystyle \frac{1}{1-2p}\, N   }       &  \qquad p<\tfrac{1}{2},\\[0.3cm]
{\displaystyle N\,\ln N     }     &  \qquad p=\tfrac{1}{2},\\[0.2cm]
{\displaystyle A(p)\,N^{2p}  }    &  \qquad \tfrac{1}{2}<p\leq 1,
\end{cases} 
\end{equation}
with 
\begin{equation}
\label{A}
A(p)=\frac{1}{(2p-1)\,\Gamma(1+2p)}\,,
\end{equation}
\end{subequations}
where $\Gamma(\cdot)$ is the Euler gamma function.

Equation \eqref{LN-sol} shows that as the copying probability $p$ is varied,
there is a transition from sparse regimes arising when $p<\frac{1}{2}$ to dense regimes when
$p\geq \frac{1}{2}$.  The average degree remains finite as $N\to\infty$ in sparse regimes and diverges in dense regimes---logarithmically with $N$ at the
transition point $p=\frac{1}{2}$ and algebraically for $p>\frac{1}{2}$.  The change in the
dependence of $L(N)$ and many other network properties as a function of $p$
is a major feature of the copying model.

Parenthetically, we can obtain the asymptotics of~\eqref{LN-sol}, with the
exception of the amplitude $A(p)$, by considering the continuum limit of
\eqref{LN-exact}.  In this limit, we treat $N$ as a continuous variable and
recast the exact difference equation \eqref{LN-exact} into the differential
equation
\begin{equation}
\label{LN}
\frac{dL(N)}{dN}=1+2p\,\frac{L(N)}{N}\,,
\end{equation}
whose solution recovers the exact asymptotics given by~\eqref{LN-exact} for
$p\leq \frac{1}{2}$.  In this range, the leading asymptotics are independent of $L(0)$; the
initial condition plays no role.  For $p>\frac{1}{2}$, the continuum
solution has the correct $N$ dependence, $L\sim N^{2p}$, but the amplitude
depends on $L(1)$.  The replacement of \eqref{LN-exact} by
\eqref{LN} is accurate only when $N\gg 1$. The dependence 
on the initial condition indicates that the behavior at small $N$ affects the outcome and hence 
the continuum approach cannot be trusted whenever there is 
 the dependence on $L(1)$.

Logarithmic and power-law densifications given in~\eqref{LN-sol} have been
observed in citation graphs, the autonomous systems graph, software networks,
and other social and information networks~\cite{GNC,sergi,kleinberg,kleinberg2}.
Network densification also occurs in models that are based on accelerated
network growth~\cite{accelerated1,accelerated-BU,accelerated2,accelerated3}.
In these models, densification arises by introducing a time-dependent attractiveness to
the nodes.  Our approach is fundamentally distinct, as densification is an
emergent property of the dynamics.

\subsection{The variance $V(N)$}
\label{variance}

We now study the variance $V(N)\equiv \langle L_N^2 \rangle-\langle L_N\rangle^2$, which characterizes
the fluctuations in the random variable $L_N$.  This variance exhibits a
richer dependence on $N$ than the average number of links $L(N)$, with a new
transition at $p=\frac{1}{4}$, in addition to the transition at $p=\frac{1}{2}$.  To
determine the variance, we need to consider the copying process in more
detail.  When a new node attaches to a randomly selected target node of
degree $k$, it also attaches to $a$ of its neighbors by copying, where $a$ is
a random variable that can range from 0 to $k$.  Thus the number of links
changes according to
\begin{equation}
\label{L_N}
L_{N+1}=L_N+1+a.
\end{equation}
Since connections to each of the neighbors of the target occurs independently
with probability $p$, the probability $Q(a|k)$ that $a$ additional links
are made to the neighbors of a target of degree $k$ is
\begin{equation}
\label{Pi}
Q(a|k)=\binom{k}{a}\, p^a (1-p)^{k-a}.
\end{equation}

Averaging \eqref{L_N} we obtain
\begin{equation}
\label{LN-av}
L(N+1)=L(N)+1+\langle\overline{a}_k\rangle.
\end{equation}
Here $\overline{a}$ denotes the average over all possible values of $a$ for a
target node of degree $k$, and $\langle\ldots\rangle$ denotes the average
over all target nodes and hence over all possible degrees.  Using \eqref{Pi}
we compute
\begin{equation}
\label{a-av}
\overline{a}=\sum_{a=0}^k aQ(a|k)=pk\,,
\end{equation}
and thus \eqref{LN-av} reduces to \eqref{LN-exact}, as it must.

We now extend this approach to compute the variance.  Squaring
Eq.~\eqref{L_N} gives
\begin{equation*}
L_{N+1}^2=L_N^2+1+a^2+2L_N+2a+2L_Na\,,
\end{equation*}
which, after averaging, becomes
\begin{equation*}
\langle L_{N+1}^2\rangle = \langle L_N^2\rangle +1+\big\langle\overline{a^2} \big\rangle
+2L(N)+2\langle\overline{a}\rangle + 2\langle L_N\overline{a}\rangle.
\end{equation*}
To compute $\langle\overline{a^2}\rangle$ and
$2\langle L_N\overline{a}\rangle$, we use \eqref{Pi} to obtain
\begin{equation}
\label{a2-av}
\overline{a^2}=\sum_{a=0}^k a^2Q(a|k)=p^2k^2+p(1-p)k\,.
\end{equation}
Therefore
\begin{equation}
\label{a2-av-av}
\big\langle\overline{a^2} \big\rangle=p^2 \langle k^2\rangle+p(1-p)\langle k\rangle.
\end{equation}
Further
\begin{equation}
\label{avLNa}
2\langle L_N\overline{a}\rangle=2p\langle L_N k\rangle=\frac{2p}{N}
\left\langle L_N \sum k\right\rangle\,,
\end{equation}
where the sum is over all $N$ nodes of the network.  Since $\sum k=2L_N$ we
conclude that
\begin{equation}
\label{LNa}
2\langle L_N\overline{a}\rangle = \frac{4p}{N}\,\langle L_N^2\rangle
\end{equation}
Using \eqref{a2-av-av}--\eqref{LNa} and $\langle k\rangle=2L(N)/N$ we find
\begin{eqnarray*}
\langle L_{N+1}^2\rangle  &={\displaystyle \Big( 1+\frac{4p}{N}\Big) \langle L_N^2\rangle
                           +1+p^2 \langle k^2\rangle} \\
& \hskip 1cm {\displaystyle +2 \Big( 1+\frac{3p-p^2}{N}\Big)\, L(N)}\,.
\end{eqnarray*}
Subtracting the square of \eqref{LN-exact} from this equation, we thereby
find that the variance evolves according to
\begin{eqnarray}
\label{VN}
V(N\!+\!1)& = {\displaystyle \Big(1+\frac{4p}{N}\Big)V(N)
+2p(1-p)\,\frac{L(N)}{N}}\nonumber\\
&{\displaystyle -\frac{4p^2}{N^2}\,L(N)^2+p^2 \langle k^2\rangle}.
\end{eqnarray}

Equation \eqref{VN} is exact but not closed as it contains
$\langle k^2\rangle$.  To close \eqref{VN} we need to express
$\langle k^2\rangle$ as a function of $L(N)$ and $V(N)$.  We have not found
such an expression and its existence seems doubtful.  To make progress, we
first estimate the asymptotic behavior of \eqref{VN} using arguments that
should apply asymptotically.  As long as we are merely interested in the
dependence of $V(N)$ on $N$ and not on amplitudes, we can replace \eqref{VN}
by the differential equation
\begin{eqnarray}
\label{VN:DE}
\frac{dV(N)}{dN}& = {\displaystyle \frac{4p}{N}\,V(N)
+2p(1-p)\,\frac{L(N)}{N}}\nonumber\\
&{\displaystyle -\frac{4p^2}{N^2}\,L(N)^2+p^2 \langle k^2\rangle}.
\end{eqnarray}

The first term on the right leads to superlinear growth, $V\sim N^{4p}$, when
$p>\frac{1}{4}$ and linear growth for $p<\frac{1}{4}$.  At $p=\frac{1}{4}$,
Eq.~\eqref{VN:DE} becomes $\frac{dV}{dN}= \frac{V}{N}+\text{const}$; hence
the variance acquires an additional logarithmic correction: $V\sim N\ln N$.
To summarize, we anticipate that the asymptotic behavior of the variance is
given by
\begin{equation}
\label{VN-sol}
V(N) \sim  
\begin{cases}
{\displaystyle N        }       &  \qquad p<\tfrac{1}{4},\\[0.1cm]
{\displaystyle N\,\ln N  }    &  \qquad p=\tfrac{1}{4},\\[0.1cm]
{\displaystyle N^{4p}     }  &  \qquad \tfrac{1}{4}<p<1\,.
\end{cases} 
\end{equation}

To derive $V(N)$ in the regime $p>\frac{1}{4}$ in a more principled way, we
need $\langle k^2\rangle$, as mentioned above.  To derive
$\langle k^2\rangle$ requires information about the degree distribution that
will be discussed in Sec.~\ref{degree}.  Here we merely quote the pertinent
results that will be used to derive of $V(N)$.  In the range
$p<p_2=\sqrt{2}-1$, the second moment is given by Eq.~\eqref{kk-av} in the
next section.  Using this result in Eq.~\eqref{VN:DE}, the evolution of the
variance is given by
\begin{equation}
\label{VN-exact}
\frac{dV(N)}{dN}=\frac{4p}{N}\,V(N) +B(p)\,,
\end{equation}
with 
\begin{equation}
\label{Bp}
B(p)=\frac{2p(1-5p+2p^2)}{(1-2p)^2}+
\frac{2p^2}{1\!-\!2p}\,\frac{3\!+\!2p\!-\!p^2}{1\!-\!2p\!-\!p^2}\,.
\end{equation}
As long as $p<p_2$, the rational function $B(p)$ is finite and positive. Solving
\eqref{VN-exact} gives, for $p<p_2$,
\begin{equation}
\label{VN-exact-sol}
V(N) =  
\begin{cases}
(1-4p)^{-1}B(p)\,N  &  \qquad p<\tfrac{1}{4},\\[0.15cm]
B(1/4)\,N\ln N         &  \qquad p=\tfrac{1}{4},\\[0.15cm]
\sim N^{4p}            &  \qquad p>\tfrac{1}{4}\,.
\end{cases} 
\end{equation}
These results improve on \eqref{VN-sol} because \eqref{VN-exact-sol} gives
the amplitude in the range $p\leq \frac{1}{4}$.  For $p>\frac{1}{4}$, the
amplitude cannot be computed within a continuum approach.

The behavior \eqref{VN-exact-sol} is established  for $p<p_2$, but 
we can extend the $V(N)\sim N^{4p}$ asymptotic to the $p>p_2$ range by noticing that the second,
third, and fourth terms on the right-hand side of \eqref{VN:DE} are of order
$\text{max}[1, N^{2p-1}]$, $\text{max}[1,N^{4p-2}]$, and $N^{p^2+2p-1}$, respectively.  [The last result
follows from Eq.~\eqref{mj}.] These terms are all subdominant with respect
to the first term on the right, which is of order $N^{4p-1}$.  Thus we
conclude that $V(N)\sim N^{4p}$ for all $p>\frac{1}{4}$.

The above results for the number of links and its variance lead us to the
following conclusions:
\begin{enumerate}
\item When $p<\tfrac{1}{4}$, the variance $V(N)$ grows linearly with
  $N$. Fluctuations are asymptotically negligible because
  $\sqrt{V(N)}/L(N)\to 0$ as $N\to\infty$. Thus we anticipate that the
  distribution $P(L,N)$ of the number of links may be asymptotically Gaussian
  when $p<\tfrac{1}{4}$.

\item The variance scales as $N^{4p}$ when $p>\tfrac{1}{4}$, thereby
  suggesting that the distribution $P(L,N)$ is non-Gaussian when
  $p>\tfrac{1}{4}$.

\item In the dense phase ($p>\tfrac{1}{2}$) the magnitude of fluctuations is
  the same as the average: $\sqrt{V(N)} \sim L(N)$.  
\end{enumerate}

The last point implies that the number of links does {\em not} self-average.
This feature leads to a wide diversity between individual realizations of the
network.  In particular, the first few steps of the network growth are
crucial to shaping its asymptotic evolution.

\section{Degree Distribution}
\label{degree} 

We now study the degree distribution, both because of its fundamental nature
in characterizing the network and because the second moment of this
distribution is an essential ingredient in the variance $V(N)$ from the
previous section.  We will argue that the copying model leads to dramatically
different degree distributions in the sparse ($p<\frac{1}{2}$) and dense
($p\geq \frac{1}{2}$) regimes.  In the sparse regime, the degree distribution has an
algebraic tail and we can also write the number of nodes of degree $k$ in the
scaling form $N_k=Nn_k$, which simplifies the analysis.  In the dense regime,
the degree distribution is anomalous in nearly all respects and thus far
defies a complete analytical description.  Finally, we will use the second
moment of the degree distribution to provide a more complete derivation of
$V(N)$.

In what follows, we assume that $N$ is sufficiently large that we can employ
a continuum approach.  Let $N_k(N)$ be the number of nodes of degree $k$ in a
network of $N$ nodes.  The degree distribution evolves according to
\begin{subequations}
\label{NkN-mk}
\begin{equation}
\label{NkN}
\frac{dN_k}{dN} = \frac{N_{k-1}-N_k}{N}+p\,\frac{(k-1)N_{k-1}-kN_k}{N}+m_k.
\end{equation}
The first two terms on the right account for the contributions due to
attachment to a randomly selected target node, the next two terms account for
attachment to the neighbors of the target node, and the last term
\begin{equation}
\label{mk}
m_k=\sum_{s\geq k-1} n_s
\binom{s}{k\!-\!1} p^{k-1} (1-p)^{s-k+1}
\end{equation}
\end{subequations}
is the probability that the new node acquires a degree $k$.  Each term in the
above sum accounts for the contribution due to a target node of degree $s$ in
which the new node attaches to the $k-1$ neighbors of this target.  Here
$n_s\equiv N_s/N$ denotes the fraction of nodes of degree $s$.

Notice that the rate equations \eqref{NkN} satisfy two basic sum rules:
$\sum_k n_k=1$, i.e., the network contains $N$ nodes, and the value of
$\sum_k kn_k$ is consistent with the total number of links growing according
to \eqref{LN}.  The first sum rule is verified by summing Eq.~\eqref{NkN}
over all $k\geq 1$.  The first four terms on the right trivially give zero.
For the last term, we use $\sum_{s\geq 1} n_s= 1$ and the binomial identity,
$\sum_{0\leq a\leq s} \binom{s}{a} p^{a} (1-p)^{s-a}=1$, to conclude that
$\sum_{k\geq 1}m_k=1$, thus giving $\sum_k n_k=1$.  In a similar spirit,
multiplying \eqref{NkN} by $k$ and summing over $k\geq 1$ gives \eqref{LN}.

\subsection{Sparse regime}
\label{sparse}

In the sparse regime, we make the standard assumption~\cite{GNR} that the
fractions $n_k=N_k/N$ are independent of $N$ for $N\gg 1$.  With this ansatz,
we recast Eq.~\eqref{NkN-mk} as
\begin{align}
\label{nk-long}
[2+p(k+1)]n_{k+1}&=[1+pk]n_k\nonumber\\
&\hskip 0.5cm +\sum_{s\geq k} n_s \binom{s}{k} p^{k} (1-p)^{s-k}\,.
\end{align}
While this is \emph{not} a recurrence, we can use this equation to determine
the behavior of low-order moments of the degree distribution. For instance, 
multiplying \eqref{nk-long} by $k(k+1)$ and summing over all $k\geq 0$ gives,
after some straightforward steps,
\begin{equation}
\label{kk-av}
\langle k^2\rangle=\sum_{k\geq 1} k^2n_k=\frac{2}{1-2p}\,\frac{3+2p-p^2}{1-2p-p^2}\,.
\end{equation}
Thus $\langle k^2\rangle$ is finite for $p<p_2$, where $p_2=\sqrt{2}-1$ is
the positive root of the polynomial $1-2p-p^2=0$. We used \eqref{kk-av} 
in deriving  \eqref{VN-exact}--\eqref{Bp} and establishing \eqref{VN-exact-sol} for $p\leq \frac{1}{4}$. 
We also note that \eqref{kk-av}
reduces to $\langle k^2\rangle=6$ for $p=0$. This last result can be verified
by recalling that the copying model reduces to random recursive trees when
$p=0$, and the degree distribution for random recursive trees is
$n_k=2^{-k}$.

To extract the asymptotics of $n_k$ from Eq.~\eqref{nk-long}, first notice
that for large $k$, the summand on the right is sharply peaked around
$s\approx k/p$ and thus reduces to~\cite{korea,protein}
\begin{equation*}
n_{k/p}\sum_{s\geq k} 
\binom{s}{k} p^{k} (1-p)^{s-k}=p^{-1}n_{k/p}\,,
\end{equation*}
where we use a binomial identity~\cite{knuth} to compute the sum.  Thus the
equation for the degree distribution reduces to
\begin{equation}
\label{nk}
[2+p(k+1)]n_{k+1}=[1+pk]n_k+p^{-1}n_{k/p}\,.
\end{equation}
This is now a non-local recurrence, as the value of $n_{k+1}$ depends both on
$n_k$ and $n_{k/p}$, where the index $k/p$ is generally much larger than $k$
itself.

\begin{figure}
\subfigure[]{\includegraphics[width=0.4\textwidth]{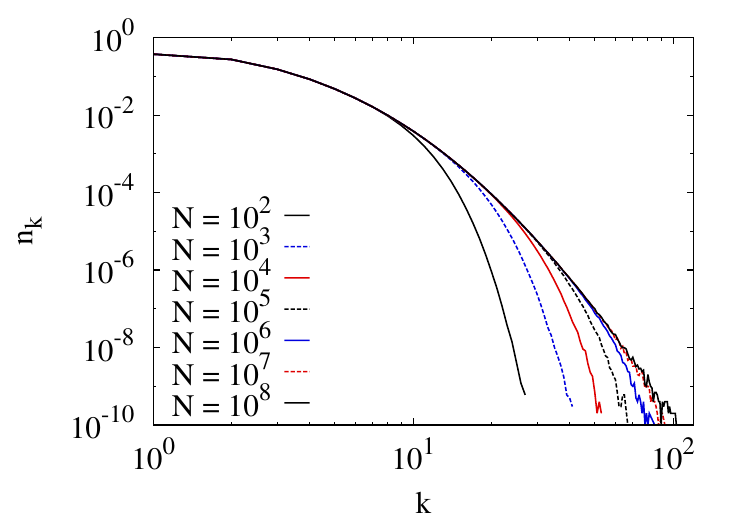}}
\subfigure[]{\includegraphics[width=0.4\textwidth]{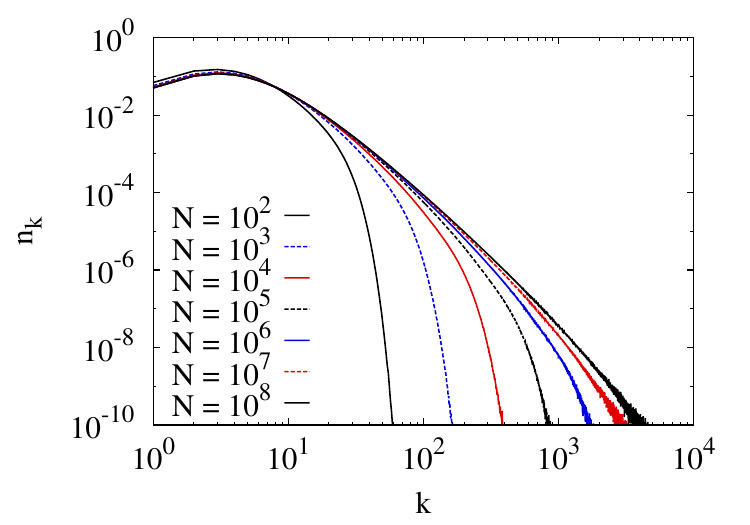}}
\caption{The scaled degree distributions in the sparse regime for (a) $p=0.1$
  and (b) $p=0.4$, For each $N$, the number of realizations is $10^{10}/N$.
}
\label{fig:deg-dist-sparse}
\end{figure}

While we have not found a systematic way to solve such a recurrence, we make
the assumption (justifiable a posteriori) that $n_k$ decays slower than
exponentially in $k$.  This allows us to replace differences by derivatives
in \eqref{nk} to give
\begin{equation}
\label{nk-eq}
\frac{d}{dk}\,[1+pk]n_k=p^{-1}\,n_{k/p} - n_k\,.
\end{equation}
This ordinary differential equation is still non-local, but it nevertheless
admits the algebraic solution $n_k\sim k^{-\gamma}$ for $k\gg 1$.
Substituting this ansatz in~\eqref{nk-eq} gives the following transcendental
relation for the degree distribution exponent
\begin{equation}
\label{gamma}
\gamma=1+p^{-1}-p^{\gamma-2}\,.
\end{equation}

Equation \eqref{gamma} has two solutions in the $(\gamma,p)$ plane.  One,
$\gamma=1$, is unphysical because it violates the sum rule
$\sum_{k\geq 1}n_k=1$.  The other applies for $0\leq p<\frac{1}{2}$.  In this case,
the exponent $\gamma=\gamma(p)$ is a monotonically decreasing function of
$p$, with $\gamma(0)=\infty$ and $\gamma(\frac{1}{2})=2$.  The feature that $\gamma$
is always greater than 2 is consistent with the sparseness of the network, in
which $\langle k\rangle =\sum_{k\geq 1}kn_k$ is finite.

Numerical results for the degree distribution in the sparse regime show that
for small $k$, the $n_k$ quickly converge to a stationary limit as a function
of $N$ (Fig.~\ref{fig:deg-dist-sparse}).  For larger $k$, the degree
distribution slowly converges to a power-law asymptotic tail whose exponent
is consistent with the prediction given in \eqref{gamma}.  This convergence
becomes progressively slower as $p$ approaches $\frac{1}{2}$.  This slow approach to
the asymptotic behavior was previously observed in a related model for
protein interaction networks~\cite{korea}, and seems to stem from the
non-locality of the equation for the degree distribution.

\subsection{Dense regime}
\label{densifying}

\begin{figure}[ht]
\subfigure[]{\includegraphics[width=0.4\textwidth]{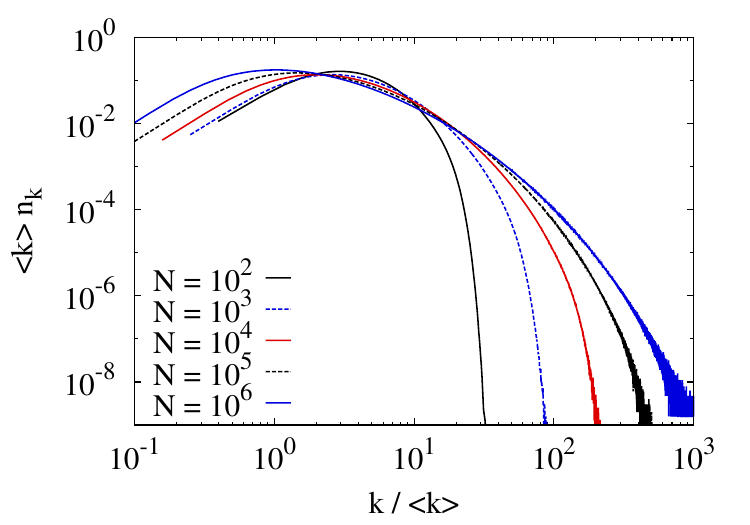}}
\subfigure[]{\includegraphics[width=0.4\textwidth]{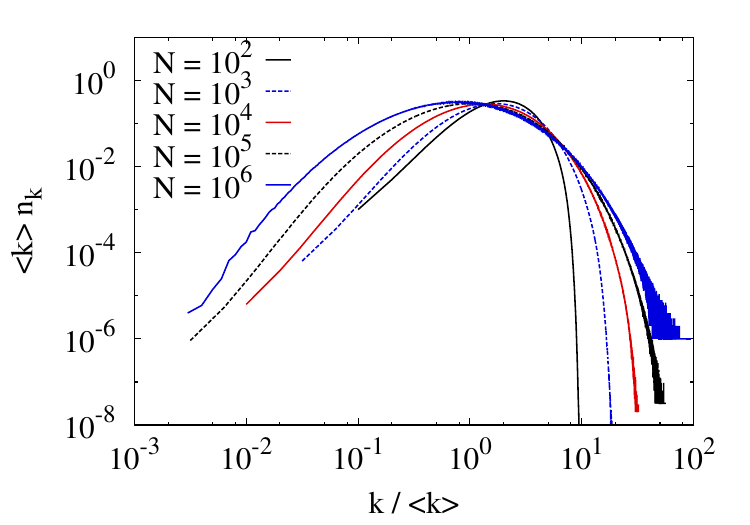}}
\subfigure[]{\includegraphics[width=0.4\textwidth]{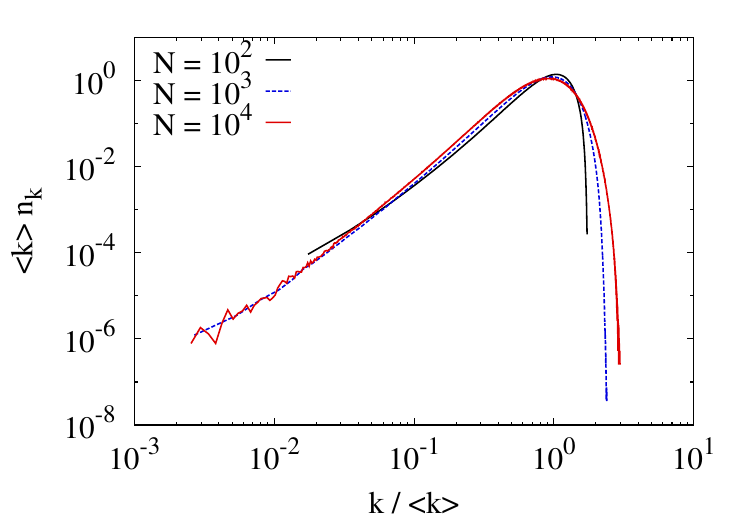}}
\caption{The degree distributions in the dense regime for: (a) $p=0.6$, (b)
  $p=0.75$ and (c) $p=0.9$.  For each $N$, the number of realizations is
  $10^{10}/N$.  }
\label{fig:deg-dist-dense}
\end{figure}

The degree distribution has a very different nature in the dense regime.
Instead of a power-law tail, the degree distribution has a well-defined peak
(Fig.~\ref{fig:deg-dist-dense}) whose location is determined by the mean
degree, which grows as $N^{2p-1}$, see Eq.~\eqref{LN-sol}.  An important
feature of the degree distribution in the dense regime is that the fractions
of nodes of degree $k$, $n_k$, are no longer stationary.  To show that the
distribution is not a power law as well as the lack of stationarity, let us
assume the converse and derive a contradiction.  We thus assume that
$n_k\sim k^{-\gamma}$ and that $n_k$ is independent of $N$.  Using this form
for $n_k$, the number of links in a finite network is given by
\begin{equation}
\label{estimate}
L=\frac{N}{2}\,\langle k\rangle=\frac{N}{2}\sum_{k=1}^{k_{\rm max}} kn_k\sim
  Nk_{\rm max}^{2-\gamma}\,,
\end{equation}
where $k_{\rm max}$ denotes the largest expected degree in a network of $N$
nodes.  We estimate this maximal degree by the standard extremal condition
(see, e.g., \cite{G87}) $N\sum_{k\geq k_{\rm max}} n_k=1$; namely, that there
is of the order of a single node whose degree is $k_{\rm max}$ or greater.
This relation gives $k_{\rm max}\sim N^{1/(\gamma-1)}$, so that
\eqref{estimate} reduces to
\begin{equation}
\label{est}
L\sim Nk_{\rm max}^{2-\gamma}\sim N^{1/(\gamma-1)}\,.
\end{equation}
On the other hand, Eq.~\eqref{LN} gives $L\sim N^{2p}$.  These two results
are consistent only when $2p(\gamma-1)=1$, and this consistency condition
agrees with \eqref{gamma} only at $p=\frac{1}{2}$.  Thus we conclude that for
$p>\frac{1}{2}$, the degree densities $n_k$ must depend on $N$, and further, that the
degree distribution is not algebraic in $k$.

\begin{figure}
\includegraphics[width=0.4\textwidth]{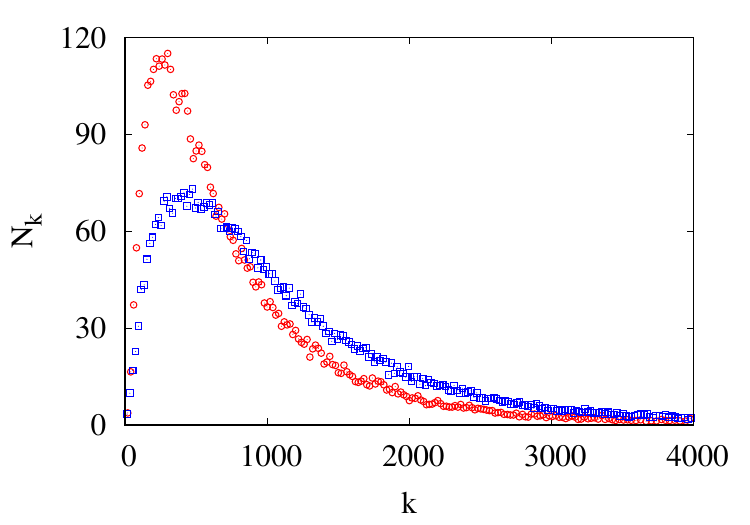}
\caption{The degree distributions for two representative realizations of the
  copying model for $p=0.75$ for a network of $N=10^5$ nodes.  The data are
  averaged over a 20-point range. }
\label{fig:non-self}
\end{figure}

Because of the non-locality of Eq.~\eqref{NkN-mk} and the non-stationary
nature of the solution, we have not found an analytical solution for the
degree distribution in the dense regime.  We therefore report on simulation
results.  Figure~\ref{fig:deg-dist-dense} shows the degree distribution,
averaged over many realizations, for representative values of $p$, with $N$
ranging between $10^2$ and $10^6$.  For each $N$, the number of realizations
is $10^{10}/N$.  These data clearly show that the degree densities are not
stationary and that scaling the degree by the average degree
$\langle k\rangle$ does not collapse the data onto a single universal curve
for networks with $10^6$ nodes or less.  It is also worth noting that the
degree distributions all exhibit a single peak, so that nodes of small
degrees do not exist for $N\to\infty$.  This behavior contrasts sharply with
the sparse regime where the degree distribution is dominated by the
smallest-degree nodes.  Finally, the degree distribution is non
self-averaging in the dense regime, as there is a wide disparity in the
degree distributions of individual network realizations
(Fig.~\ref{fig:non-self}).

\section{Cliques and Other Motifs}
\label{sec:cliques}

As $p$ is increased, it becomes increasingly likely that triangles are
generated when each node is introduced.  With this increased frequency for
triangles, there is a concomitant increased propensity for the appearance of
$m$-cliques --- complete subgraphs of $m$ nodes.  To investigate this
feature, we extend the approach of Sec.~\ref{sec:links} for the number of
links, to first account for the average number of triangles, and then the
average number of $m$-cliques for general $m$.

\subsection{Triangles}
\label{triangle}

We begin by giving a (trivial) lower bound for the number of triangles $T_N$
in a network of $N$ nodes.  If there was no copying, the number of links
$L_N$ would equal $N-1$ in the resulting tree network, so that no triangles
would exist.  For each copying event, the number of links increases by 1
while the number of nodes remains fixed, and at least one triangle is
created.  This reasoning gives the bound
\begin{equation}
\label{lowerbound}
T_{N} \geq L_N - (N-1)\,.
\end{equation}
For $p<\frac{1}{2}$, this bound, together with \eqref{LN-sol}, gives, for the average
number of triangles, 
\begin{equation*}
T(N) \equiv \langle T_N\rangle \geq \frac{2pN}{1-2p}\,.
\end{equation*}
We will see that the average number of triangles grows linearly with $N$ when
$p<\frac{1}{2}$, while for $p>\frac{1}{2}$, the growth of $T(N)$ is superlinear in $N$.

\begin{figure}[h]
\begin{center}
\includegraphics[width=0.2\textwidth]{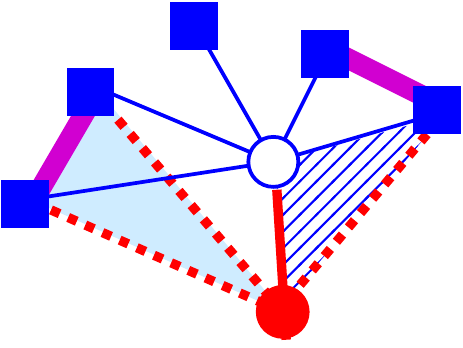}
\caption{\small Counting triangles.  The target node (open circle) has five
  neighbors (squares), two of which are joined by `clustering' links (heavy
  lines).  When a new node (filled circle) is introduced, three copying links
  (dashed) create three new triangles (one is hatched for illustration) and
  one new triangle by induced linking (shaded).}
\label{TN}
\end{center}
\end{figure}

In each successful copying event a triangle is generated that consists of the
new node, the target node and the neighbor that receives a copied link.  We
term this triangle-generating mechanism as \emph{direct linking}.  If links
to two neighbors of the target are created, then two triangles necessarily
arise by direct linking.  Additional triangles may be created by a process
that we term \emph{induced linking}: when links to two neighbors of the
target are created and these neighbors were previously linked, then a third
triangle is created (shaded in Fig.~\ref{TN}).

To determine $T(N)$, we need to account for both of these mechanisms.
Suppose that the target node has degree $k$ and that its neighbors are
connected via $c$ `clustering' links (Fig.~\ref{TN}).  If $a$ links to the
neighbors are made by copying, the number of triangles increases on average
by
\begin{equation}
\label{deltaT}
\Delta T = a + \frac{a (a-1)}{2} \frac{c}{k (k-1)/2}\,.
\end{equation}
The first term on the right accounts for direct linking and the second for
induced linking.  For the latter, we count how many of $a(a-1)/2$ possible
links between $a$ neighbors of the target, which also connect to the new
node, are actually present.  We now average \eqref{deltaT} with respect to
the binomial distribution \eqref{Pi} for $a$.  This elementary calculation,
together with the already-known result $\overline{a}=pk$ from
Eq.~\eqref{a-av}, gives
\begin{equation*}
\overline{a(a-1)}=\sum_{a=0}^k a (a-1) Q(a|k)=p^2 k (k-1),
\end{equation*}
from which we obtain the compact result
\begin{equation}
\label{deltaT2}
\overline{\Delta T} =  pk + p^2 c.
\end{equation}
The term $p^2 c$ in Eq.~\eqref{deltaT2} can be understood by noting that two
previously connected neighbors also get connected to the new node with
probability $p^2$ since linking to each node occurs independently.

We now express the average degree $\langle k \rangle$ via $L(N)$ and the
average number of clustering links $\langle c \rangle$ via $T(N)$.  The
former relation is known, while to determine the latter we note that $c$
equals the number of triangles that contain the target node. Thus
\begin{equation}
\label{relationsT}
\langle k \rangle = \frac{2 L(N)}{N}\,,\qquad \langle c \rangle = \frac{3 T(N)}{N}\,.
\end{equation}
Using \eqref{relationsT}, we average the increment of the number of triangles
in \eqref{deltaT2} to obtain
$\langle \overline{\Delta T} \rangle = 2 p L/N + 3 p^2 T/N$ each time a new
node is added.  Therefore the number of triangles evolves according to
\begin{equation}
\label{tevolution}
T(N\!+\!1) =\left( 1 + \frac{3 p^2}{N} \right) T(N) + 2p \frac{L(N)}{N}\,.
\end{equation}
Solving this recurrence equation (see Appendix~\ref{app:TN}) gives the
asymptotic behaviors
\begin{subequations}
\label{T-tot}
\begin{equation}
\label{TN-sol}
T(N)= 
\begin{cases}
\displaystyle{\frac{2p}{(1\!-\!2p)\,(1\!-\!3p^2)}\,\, N}      &  \quad p< \frac{1}{2}\,,\\[0.35cm]
4 N\,\ln N                                    &  \quad p=\frac{1}{2}\,,\\[0.2cm]
\displaystyle{\frac{A(p)}{1\!-\!3p/2} \,N^{2p} }     &  \quad \frac{1}{2}<p < \frac{2}{3}\,,\\[0.4cm]
\displaystyle{\tfrac{4}{3}A\!\left(\tfrac{2}{3}\right) N^{4/3} \ln N  }   &  \quad p=\frac{2}{3}\,,\\[0.25cm]
C(p)\,N^{3p^2}      &  \quad \frac{2}{3}<p\leq 1,
\end{cases} 
\end{equation}
with $A(p)$ given by \eqref{A} and
\begin{equation}
\label{C}
C(p)=\frac{2}{(3p-2)\,(3p^2-1)\,\Gamma(3 p^2 +1)}\,.
\end{equation}
\end{subequations} 
Notice that for $N\gg 1$, the recursion \eqref{tevolution} reduces to the
differential equation
\begin{equation*}
\frac{dT}{dN} = 3 p^2 \frac{T}{N}  + 2p \frac{L}{N}\,,
\end{equation*}
whose solution coincides with \eqref{TN-sol}, except for the amplitude in the
regime $p>\tfrac{2}{3}$, which cannot be determined within the continuum approach.

Equation~\eqref{TN-sol} exhibits several striking features.  First, the
triangle density (the average number of triangles per node) converges to a
non-vanishing value for all $0 < p<\frac{1}{2}$, as observed in many
empirical complex networks.  This linearity arises because for any $p>0$ a
non-zero number of triangles are typically created when each node is added.
Second, the average number of triangles $T(N)$ undergoes phase transitions at
$p=\frac{1}{2}$ and at $p=\tfrac{2}{3}$.  Although there is change in the $N$
dependence at $p=\frac{1}{2}$, the average number of triangles continues to
scale linearly with the number of links for any $p<\tfrac{2}{3}$.  However,
beyond $p=\tfrac{2}{3}$, the number of triangles grows \emph{faster} than the
number of links.

\subsection{Cliques}
\label{kC}

We can extend the above considerations to treat complete subgraphs, or
motifs, of arbitrary size $m$ (with links and triangles corresponding to
motifs of size 2 and 3 respectively).  Let $K_m(N)$ be the average number of
such motifs in a network of $N$ nodes, with $K_2(N) \equiv L(N)$ and
$K_3(N) \equiv T(N)$.

To determine the number of quartets---cliques of size four---we use similar
reasoning that led to Eq.~\eqref{deltaT2}.  We thus find that adding a node
gives, for the average increase $\overline{\Delta K_4}$ in the number of
quartets:
\begin{equation}
\label{deltaQ}
\overline{\Delta K_4} =  p^2 c + p^3 d.
\end{equation}
Here $d$ is the number of triangles whose vertices are all neighbors of the
target node.  Using $\langle c \rangle = 3T /N$ and
$\langle d \rangle = 4K_4/N $, we find that in the large-$N$ limit the
average number of quartets evolves according to
\begin{equation}
\label{cqevolution}
\frac{dK_4}{dN} = 3 p^2 \frac{T}{N}  + 4 p^3 \frac{K_4}{N}\,,
\end{equation}
whose solution is
\begin{subequations}
\begin{equation}
\label{QN-sol}
K_4(N) \sim
\begin{cases}
N                     &   \quad 0 < p < \tfrac{1}{2},\\[0.1cm]
N^{2p}            &   \quad \tfrac{1}{2}< p < \tfrac{2}{3},\\[0.1cm]
N^{3 p^2}       &  \quad \tfrac{2}{3} <p < \tfrac{3}{4},\\[0.1cm]
N^{4 p^3}       &  \quad \tfrac{3}{4} <p \leq 1.
\end{cases} 
\end{equation}
At the transition points $p = \frac{1}{2}$, $\tfrac{2}{3}$, and
$\tfrac{3}{4}$, the corresponding algebraic factor is multiplied by $\ln N$.

We can refine the above results by incorporating the exact asymptotic
behaviors about triangles from \eqref{TN-sol}, to obtain the exact amplitudes
in the range $0\leq p\leq \tfrac{3}{4}$: 
\begin{equation}
K_4(N)= 
\begin{cases}
\displaystyle{\frac{6p^3}{(1\!-\!2p)(1\!-\!3p^2)(1\!-\!4p^3)}\,\, N} & p<\tfrac{1}{2},\\[0.3cm]
\displaystyle{6 N\,\ln N}                               &   p=\tfrac{1}{2},\\[0.1cm] 
\displaystyle{\frac{3pA(p)}{(2-3p)(1-2p^2)} \,N^{2p}} &\tfrac{1}{2}<p<\tfrac{2}{3},\\[0.4cm]   
12A\!\left(\tfrac{2}{3}\right) N^{4/3} \ln N     &  p=\tfrac{2}{3},\\[0.3cm]  
\displaystyle{\frac{C(p)}{1-4p/3}\,N^{3p^2}}     &   \tfrac{2}{3} <p < \tfrac{3}{4},\\[0.4cm]  
\tfrac{27}{16} C\!\left(\tfrac{3}{4}\right) N^{27/16} \ln N   &  p=\tfrac{3}{4},\\[0.25cm]  
\sim N^{4p^3}                                           &   \tfrac{3}{4}<p\leq 1.
\end{cases} 
\end{equation}
\end{subequations}
To obtain the amplitude in the range $\tfrac{3}{4}<p<1$ range requires an analysis of
an exact recurrence for $K_4(N)$.

More generally, the average number of cliques of $m$ nodes, $K_m(N)$,
satisfies
\begin{equation}
\label{cmevolution}
\frac{dK_m}{dN} = (m-1) p^{m-2} \frac{K_{m-1}}{N}  + m p^{m-1} \frac{K_{m}}{N}\,.
\end{equation}
Solving \eqref{cmevolution} recursively gives
\begin{subequations}
\begin{equation}
\label{Mm}
K_m =\frac{N}{1-m p^{m-1}} \prod_{j=1}^{m-2} \frac{(j+1) p^j}{1- (j+1) p^j} 
\end{equation}
in the sparse phase ($p<\frac{1}{2}$), while in the dense phase 
\begin{equation}
\label{MN-sol}
K_m \sim N^{(j+1) p^j}   \qquad     {\rm for}  \quad \frac{j}{j+1}<p < \frac{j+1}{j+2}\,,
\end{equation}
\end{subequations}
with $j = 0,1, 2,\ldots , m-1$ (the last asymptotic for $j=m-1$ holds when
$1-m^{-1} < p < 1$).  The $N$ dependence of the average number of cliques of
size $m$ therefore undergoes transitions at $p = 1 - 1/n$ with
$n = 2,\ldots,m$.  Thus the dense regime of the copying model can be
partitioned into progressively finer subintervals where there are distinct
$N$ dependences for the number of $m$-cliques.

\subsection{Star Subgraphs}

Another simple motif within a complex network is a star graph.  Part of the
reason to study star graphs is that they are simply related to the degree
distribution itself.  Let $S_j$ denote the number of star graphs with $j$
leaves (nodes of degree 1).  A node of degree $k$ is thus a central node in
$\binom{k}{j}$ subgraphs of type $S_j$.  As a consequence, the number of star
graphs and the degree distribution in a given network are related by
\begin{equation}
\label{Sj}
S_j = \sum_{k\geq j} \binom{k}{j}N_k\,.
\end{equation}
We denote by $S_j(N)$ the average number of subgraphs of type $S_j$ in a
network of $N$ nodes.  From \eqref{Sj}, there is a simple relation between
the average number of stars and the falling factorial moments of the degree
distribution:
\begin{equation}
\label{m_j}
S_j(N)= \frac{1}{j!}\,N\mu_j, \quad \mu_j=\langle  k(k-1)\ldots(k-j+1)\rangle\,.
\end{equation}
Using the evolution equation for the degree distribution, Eq.~\eqref{NkN},
the falling factorial moment, which is a function of $p$ and $N$, evolves
according to
\begin{equation}
\label{mm_j}
N\frac{d\mu_j}{dN}=(p^j+jp-1)\mu_j+j[1+(j-1)p+p^{j-1}]\mu_{j-1}\,.
\end{equation}

From \eqref{mm_j}, each factorial moment $\mu_j\equiv \mu_j(p,N)$ remains
finite, $\lim_{N\to\infty}\mu_j(p,N)\equiv\mu_j(p)$, when $p<p_j$, where
$p_j$ is the positive root of
\begin{equation}
\label{pj}
p^j+jp-1=0\,.
\end{equation}
When $p<p_j$, Eq.~\eqref{mm_j} yields the recurrence
\begin{equation*}
\mu_j(p) =  j\,\frac{1+(j-1)p+p^{j-1}}{1-jp-p^j}\,\mu_{j-1}(p)\,,
\end{equation*}
from which
\begin{equation}
\label{l_j}
\mu_j(p) =  j!\, \lambda_j(p), 
\end{equation}
where we define the shorthand notation
\begin{equation*}
\lambda_j(p)\equiv \prod_{i=1}^j \frac{1+(i-1)p+p^{i-1}}{1-ip-p^i}\,.
\end{equation*}

Generally
\begin{equation}
\label{mj}
\mu_j = 
\begin{cases}
j!\,\lambda_j(p)                    & \quad  p<p_j\,, \\[0.1cm]
j!\,\Lambda_j \ln N               &  \quad p=p_j\,,\\[0.1cm]
\sim N^{jp+p^j-1}                 & \quad p_j<p\leq 1\,,
\end{cases} 
\end{equation}
where $\Lambda_j = [1+(j-1)p_j+p_j^{j-1}]\lambda_{j-1}(p_j)$.  As a
consistency check, notice for the case $j=1$, equation \eqref{mj} reproduces the
average degree $\mu_1=\langle k \rangle=2L(N)/N$.

Combining \eqref{m_j} and \eqref{mj}, the number of stars asymptotically
behaves as
\begin{equation}
\label{numb-Sj}
S_j(N) = 
\begin{cases}
\lambda_j(p)  N                   &  \quad p<p_j\,,\\[0.1cm]
\Lambda_j N\ln N                &  \quad p=p_j\,, \\[0.1cm]
\sim N^{jp+p^j}                    &  \quad p_j<p\leq 1\,.
\end{cases} 
\end{equation}
Overall, the numbers of star subgraphs have a simpler $N$ dependence than
cliques because the former undergo a single transition for each $j$ at an
irrational value of $p_j$ whose first few values are:
\begin{equation*}
\begin{split}
&p_2 = \sqrt{2}-1 \,,  \\
&p_3 = -\big[2/(1 + \sqrt{5})\big]^{1/3} + \big[(1 + \sqrt{5})/2\big]^{1/3}\,,\\
&p_4 = \left[{\sqrt{2\sqrt{2}-1}-1}\right]/{\sqrt{2}}\,,
\end{split}
\end{equation*}
etc.  From \eqref{pj}, the asymptotic behavior of the threshold values are
given by $p_j\to {1}/{j}-1/j^{j+1}$ for $j\gg 1$.  In contrast, the phase
transition points for complete-graph motifs $K_m(N)$ are all rational and at
the same location for every $m$---only the number of transition points is
variable, with $m-1$ transition points.

\section{Clustering}
\label{sec:clustering}

For intermediate values of $p$, we have seen that the copying model gives
rise to non-trivial motifs, and we now investigate whether their appearance
corresponds to the emergence of significant network clustering, as might be
surmised visually in Fig.~\ref{vis}.  There are two popular measures of
network clustering: (i) the transitivity, or global clustering coefficient,
and (ii) the local clustering coefficient (see e.g.,~\cite{N10}).  The
transitivity $\tau_G$ for a connected, undirected, and simple (no multiple
links between two nodes) graph $G$ is defined as
\begin{equation}
\label{GCC}
\tau_G = 3\times \frac{\#(\text{triangles in $G$})}{\#(\text{twigs in $G$})}\,.
\end{equation}
Here, a twig is a node with two neighbors and thus looks like:
$\bullet$\!\rule[0.07cm]{.3cm}{0.01cm}\!$\bullet$\!\rule[0.07cm]{.3cm}{0.01cm}\!$\bullet$.
By definition, the transitivity is already averaged over all network nodes.

To define the local clustering coefficient, first consider an arbitrary node
${\bf n}$ of degree $k$ in the network.  The $k$ neighbors of ${\bf n}$ could
potentially be connected by up to $\binom{k}{2}$ edges.  The clustering
coefficient of node ${\bf n}$ is then defined as $c({\bf n})/\binom{k}{2}$, where
$c({\bf n})$ denotes the actual number of connections between the neighbors
of ${\bf n}$.  Finally, the local clustering coefficient $CC(N)$ is obtained
by averaging the node clustering coefficient over all nodes:
\begin{equation}
\label{ACC}
CC(G) = \frac{1}{N}\sum_{{\bf n}\in G} \frac{c({\bf n})}{\binom{k}{2}}\,.
\end{equation}

If $G$ is a tree, the above clustering coefficients vanish, while if $G$ is
the complete graph, both clustering coefficients equal one (which explains
the choice of the numerical factor in the definition~\eqref{GCC}).  We now
examine the dependence of the clustering coefficients on the copying
probability.  Each network realization leads to distinct values for the
clustering coefficients.  In fact, the transitivity is non-self-averaging
when $p>\frac{1}{2}$.  In this dense region, however, the transitivity vanishes as
$N\to\infty$ so that the lack of self averaging does not pose any
difficulties.  Conversely, for sufficiently small $p$, where the transitivity
is non-zero in the $N\to\infty$ limit, the transitivity is self-averaging and
is determined from
\begin{equation}
\label{TTL}
\tau(N)\equiv \langle \tau_G\rangle = \frac{3T(N)}{S_2(N)}\,,
\end{equation}
where $T(N)$ is the average number of triangles and $S_2(N)$ is the average
number of twigs.

\begin{figure}
\centerline{
\includegraphics[width=0.4\textwidth]{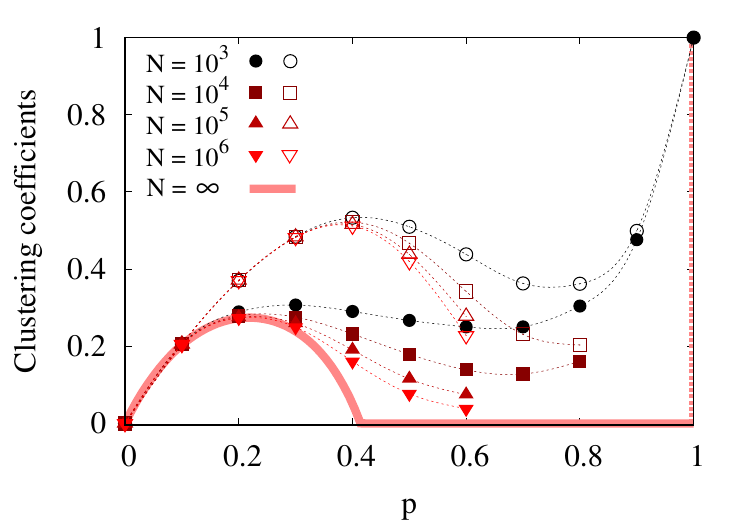}}
\caption{The transitivity $\tau(N)$ (solid symbols) and the local clustering
  coefficient $CC(N)$ (open symbols) versus the copying probability $p$ for
  networks of different sizes.  The solid smooth curve is the analytical
  expression \eqref{transitivity:inf}.  The dotted curves are guides to the
  eye. }
\label{TR_fig}
\end{figure}

To determine the transitivity in the limit $N\to\infty$, we need the average
number of triangles $T(N)$, which is given by Eq.~\eqref{TN-sol} and the
average number of twigs $S_2(N)$.  The latter is given by specializing
\eqref{numb-Sj} to $j=2$:
\begin{align}
\label{Twigs-av}
S_2(N)= 
\begin{cases}
{\displaystyle \frac{2(1+2p)}{(1\!-\!2p)(1\!-\!2p\!-\!p^2)} \,N}     &  p<p_2, \\[0.3cm]
{\displaystyle 2\,\frac{1+2p_2}{1-2p_2}\,\,N\ln N  }        &  p=p_2,\\[0.3cm]
\sim N^{2p+p^2}                                & p_2<p\leq 1.
\end{cases} 
\end{align}
With these results, the transitivity is (Fig.~\ref{TR_fig})
\begin{equation}
\label{transitivity:inf}
\tau(\infty) = 
\begin{cases}
{\displaystyle \frac{3p(1-2p-p^2)}{(1+2p)(1-3p^2)}    }   & \quad 0\leq p \leq p_2, \\[0.3cm]
0                         & \quad p_2<p<1, \\[0.3cm]
1                         & \quad p=1,
\end{cases}
\end{equation}
where $p_2=\sqrt{2}-1$ is again the positive root of the quadratic equation
$p^2+2p-1=0$.  

A perplexing feature of the transitivity is its non-monotonic dependence on
$p$, with a maximum deep in the sparse regime (at $p\approx 0.2181$). We also
emphasize that when $p_2\leq p<1$, the transitivity vanishes in the
thermodynamic limit $N\to\infty$.  However, the simulations show that even
for large networks the transitivity is non-zero and approaches zero very
slowly as $N$ increases (Fig.~\ref{TR_fig}).  This features can be understood
theoretically.  For instance, in the marginal case of $p=p_2$,
Eq.~\eqref{TTL}, in conjunction with \eqref{TN-sol} and \eqref{Twigs-av},
shows that the transitivity exhibits a slow inverse logarithmic decay:
$\tau(N)\sim (\ln N)^{-1}$.

\section{Distribution of Links and Cliques}
\label{PLN}

Because a varying number of links are added to the network each time a new
node is introduced, the \emph{distributions} of the number of links and the
number of cliques are non-trivial quantities.  Here we investigate the
asymptotic properties of these link and clique distributions by numerical
simulations, as well as basic probabilistic and extreme statistics arguments.

\subsection{Link Distribution}

\begin{figure}[ht]
\includegraphics[width=0.45\textwidth]{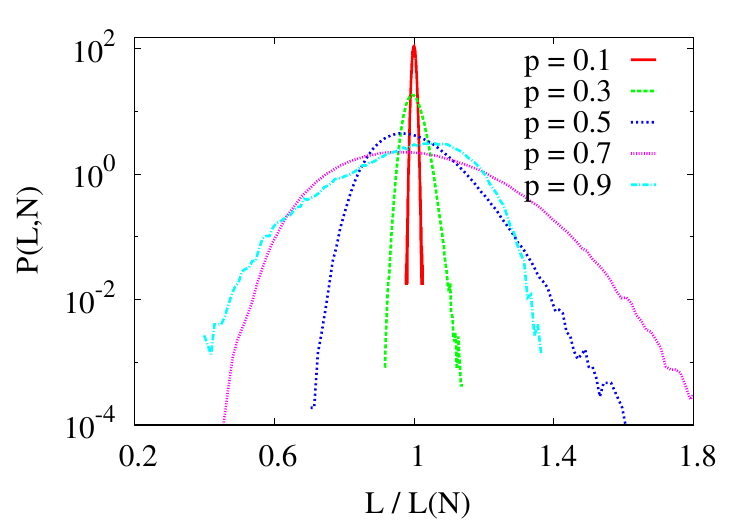}
\caption{Semi-logarithmic plot of the distribution $P(L,N)$ versus
  the scaled number of links for $N = 10^4$ and representative values of
  $p$. Data collected over $10^6$ realizations for $p$ up to $0.7$ and $10^5$
  realizations for $p=0.9$.}
\label{PLp}
\end{figure}

Let $P(L,N)$ be the probability that a network of $N$ nodes contains $L$
links: $P(L,N)=\text{Prob}(L_N=L)$. As a function of $p$, this distribution
exhibits a wide range of behaviors (Fig.~\ref{PLp}).  For $p\ll 1$, the
distribution $P(L,N)$ is visually symmetric and Gaussian in appearance.  As
$p$ is increased, $P(L,N)$ broadens considerably and is enhanced at large
argument.  Visually, $P(L,N))$ is maximally broad for $p\approx 0.7$, while
for larger $p$, the distribution progressively narrows and develops an
enhancement at small argument.

Each time a new node is introduced, the number of links increases by $1+a$,
where the random variable $a$ is the number of copying links that are created
(Eq.~\eqref{L_N}).  In the sparse phase, where the degree distribution
reaches a stationary limit with the algebraic tail $k^{-\gamma}$
(Eq.~\eqref{gamma}), the increment in the number of links $1+a$ is also drawn
from this same distribution.  For $\gamma > 3$, which occurs when
$p < \sqrt{2} - 1$, the second moment $\langle a^2 \rangle$ is finite.
Because the first two moments of the link increment are finite, one might
anticipate that the central limit theorem applies, from which
$P(L,N)$ would asymptotically be Gaussian.

However, the increments $1+a$ when each node is introduced are not
statistically independent.  A particularly fruitful copying event for a
high-degree target node increases the degrees of many neighboring nodes,
which, in turn, affects the increment in the number of links in later node
additions.  Thus the growth in the number of links is governed by a
correlated random-walk process and the central-limit theorem is not
applicable to infer the asymptotic form of $P(L,N)$.

\begin{figure}
\includegraphics[width=0.45\textwidth]{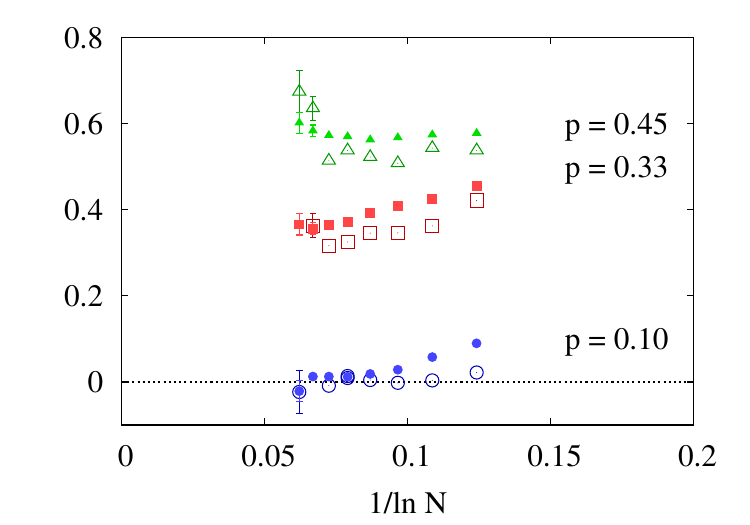}
\caption{Skewness (solid symbols) and excess kurtosis (open symbols) of the
  link distribution as a function of $N$.}
\label{skewness-kurtosis-L}
\end{figure}

From Eqs.~\eqref{LN-sol} and \eqref{VN-exact-sol}, the ratio of the square
root of the variance to the average number of links, $\sqrt{V(N)}/L(N)$
decays as $N^{-1/2}$ for $p<\frac{1}{4}$ and slower than $N^{-1/2}$ for larger $p$.
This behavior suggests that $p=\frac{1}{4}$ might be the point where $P(L,N)$ changes
in character from Gaussian to non Gaussian.  We also test the Gaussianity of
$P(L,N)$ by measuring its skewness, $\mu_3/\sigma^3$ where $\mu_n$ is the
$n^{\mathrm{th}}$ central moment and $\sigma$ is the standard deviation of
the probability distribution, and excess kurtosis, $\mu_4/\sigma^4-3$.  Both
these quantities are zero for the Gaussian distribution.  Numerically, we
find that for $p<\frac{1}{4}$, the skewness and excess kurtosis do approach zero as
$N\rightarrow\infty$, while for $p>\frac{1}{4}$, these quantities are both non-zero
as $N\rightarrow\infty$ (Fig.~\ref{skewness-kurtosis-L}).  These results
indicate that the distribution $P(L,N)$ is non Gaussian for $p>\frac{1}{4}$.

\begin{figure}
\includegraphics[width=0.4\textwidth]{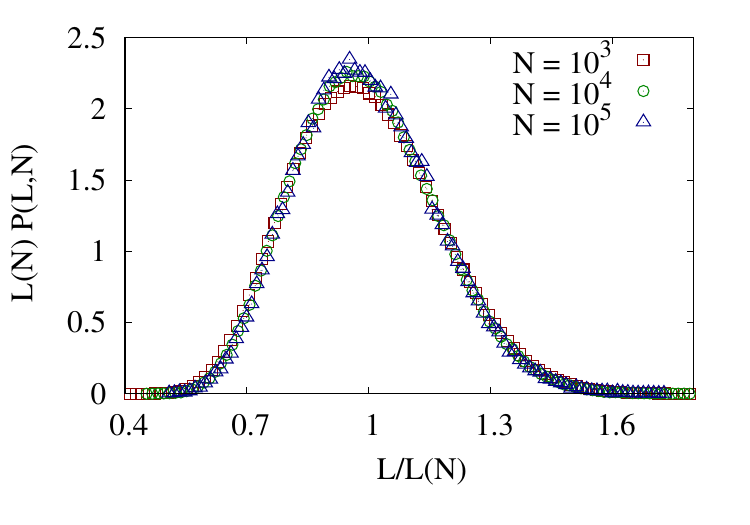}
\caption{The scaled distribution of the number of links for the copying model with
  $p=0.7$.  }
\label{finalfig}
\end{figure}

When $p>\frac{1}{2}$, the standard deviation in the number of links $\sqrt{V(N)}$
grows as $L$, and this suggests that $P(L,N)$ approaches the single-parameter
scaling form,
\begin{equation}
\label{scalingform}
P(L,N) \simeq \,\,\frac{1}{L(N)}\, \Phi(\mathsf{L})\quad \text{with} \quad \mathsf{L} = L/L(N)\,,
\end{equation}
as confirmed in Fig.~\ref{finalfig}.  In many processes that are generated by
a random-walk-like process, the scaling function $\Phi(\mathsf{L})$ has the
limiting forms~\cite{F66,G79,L88}
\begin{equation}
\label{P}
-\ln \Phi(\mathsf{L}) \sim
\begin{cases}
\mathsf{L}^{\delta_+} & \mathsf{L}\gg 1,\\[0.1cm]
(1/\mathsf{L})^{\delta_-} & \mathsf{L}\ll 1\,.
\end{cases}
\end{equation}
We now give heuristic arguments for the tail exponents $\delta_\pm$ by
considering the extreme cases where $L$ is: (i) as large as possible, and
(ii) as small as possible, and matching the distribution $P(L,N)$
in these extreme cases to the hypothesized limiting form of the full
distribution.

The maximal number of links $L_{\rm max}=N(N-1)/2$ corresponds to generating
a complete graph.  The probability $C(N)$ to construct a complete graph is
\begin{align}
\label{PiN}
C(N) = p\,\, p^2\, p^3\ldots p^{N-2} \simeq \exp\big(\tfrac{1}{2}N^2\ln p\big)\,.
\end{align}
Each factor $p^n$ gives the probability that the addition of a node to a
complete graph of $n+1$ nodes leads to a complete graph of $n+2$ nodes.  In the dense regime $L(N)= A(p)N^{2p}$
[see Eq.~\eqref{LN-sol}], so that the maximal value of the scaling variable is
$\mathsf{L}_\text{max}=L_\text{max}/L(N)\sim N^{2(1-p)}$.  Using this value of
$\mathsf{L}_{\rm max}$ and matching \eqref{P} with \eqref{PiN}, we obtain
$[N^{2(1-p)}]^{\delta_+}\sim N^2$, from which we extract the large-$\mathsf{L}$
tail exponent
\begin{equation*}
\delta_+ = \frac{1}{1-p}\,.
\end{equation*}

Conversely, the smallest possible $L$ arises if no copying connections are
made, so that the resulting network is a tree with $L=L_\text{min}=N-1$.  The
probability that no copying connections are made when a new node attaches to
a node of degree $k$ is $(1-p)^k$.  Thus the probability to generate a tree
is $(1-p)^{\sum k}$, where the sum runs over the degrees of all selected
target nodes.  The upper bound $(1-p)^{N-1}$ arises in the situation when
only leaves (nodes of degree 1) have been selected during the network
creation.  Generally one still anticipates that $\sum k \sim N$ and hence
$\ln \Phi(\mathsf{L}_\text{min})\sim N\ln(1-p)$. Since
$\mathsf{L}_\text{min}=L_\text{min}/L(N)\sim N^{1-2p}$ the matching gives
$[N^{2p-1}]^{\delta_-}\sim N$ leading to the left tail exponent
\begin{equation*}
\delta_- = \frac{1}{2p-1}\,.
\end{equation*}

To summarize, the tails of the distribution of the number of links are given
by
\begin{equation}
\ln\Phi(\mathsf{L}) \sim  - 
\begin{cases}
\mathsf{L}^{1/(1-2p)}   &  \mathsf{L}\ll 1\,,\\[0.1cm]
\mathsf{L}^{1/(1-p)}    &   \mathsf{L}\gg 1\,.
\end{cases}
\end{equation}

\subsection{Triangle Distribution}

One can also investigate the distributions of other cliques.  For triangles,
for example, the corresponding probability distribution is
$P(T,N)\equiv\text{Prob}(T_N=T)$. We make that ansatz that in the dense phase the
distribution $P(T,N)$ approaches the single-parameter scaling form,
\begin{equation}
  \label{scaling-triangles}
  P(T,N) \simeq \frac{1}{T(N)}\,\Psi(\mathsf{T})\quad \text{with} \quad \mathsf{T} = T/T(N)\,.
\end{equation}
As in the case of the link distribution, we postulate that the large-argument
tail of the scaled distribution has the form $\ln \Psi(\mathsf{T})\sim -\mathsf{T}^\delta$ for
$\mathsf{T}\gg 1$, which we expect will be valid in the dense phase $p>\frac{1}{2}$.  We now
estimate the large-argument tail of the triangle distribution by again
considering the extreme case where the number of triangles is as large as
possible.  The largest possible value of $T$ arises when a complete graph is
generated.  In this case, $T=T_\text{max}=\binom{N}{3}$ and using
Eq.~\eqref{TN-sol} the scaling variable $\mathsf{T}$ is given by
\begin{equation}
\mathsf{T}\sim  
\begin{cases}
N^{3-2p}    &      \tfrac{1}{2}<p< \tfrac{2}{3} \,, \\[0.1cm]
N^{3-3p^2}    &      p>\tfrac{2}{3} \,.
\end{cases}
\end{equation}

On the other hand, from Eq.~\eqref{PiN}, the probability to construct a
complete graph is given by $\exp(\frac{1}{2}N^2\ln p)$.  This form matches
\eqref{scaling-triangles} if the $\mathsf{T} \gg 1$ tail of the triangle
distribution is given by
\begin{equation}
\ln \Psi(\mathsf{T}) \sim  - 
\begin{cases}
  \mathsf{T}^{2/(3-2p)}     &      \tfrac{1}{2} < p < \tfrac{2}{3}\,, \\[0.1cm]
  \mathsf{T}^{2/(3-3p^2)} & \tfrac{2}{3} < p < 1\,.
\end{cases}
\end{equation}
This same line of reasoning can be straightforwardly adapted to obtain the
large-argument tail of the distribution of $m$-cliques.

\section{Second-Neighbor Connections}
\label{sec:2nd-neighbor}

Suppose that in addition to connecting to the neighbors of the target with
probability $p$, a new node \emph{also} connects to the second neighbors of
the target with probability $q$.  Such a mechanism naturally arises in social
media, such as Facebook, where we are sporadically encouraged to make
connections to friends of our friends.  The surprising outcome of
second-order linking is that the probability that the network is complete
with non-zero, albeit may be very small, for any $q>0$ probability.

To estimate this completeness probability, suppose that as the network is
complete when it contains $N$ nodes.  Then the probability that the network
remains complete when the $(N+1)^{\rm st}$ is introduced is
\begin{align}
\label{CN}
C(N) &= \sum_{k=0}^N B(N,k,p)\left[1-(1-q)^k\right]^{N-k} \,,
\end{align}
where $B(r,k,p)=\binom{r}{k} p^{k}(1-p)^{r-k}$ is the binomial probability.
The factor $B(r,k,p)$ gives the probability that there are $k$ first-neighbor
connections from the new node, while the remaining factor
$\left[1-(1\!-\!q)^k\right]^{N-k}$ gives the probability that the remaining $N-k$
nodes are linked to the new node by second-neighbor connections.

We now argue that $C(N)$ approaches to 1 sufficiently quickly as $N$
increases, so that the product of these factors converges to a non-zero
value.  In the large-$N$ limit, the binomial factor becomes a Gaussian
distribution that is sharply peaked about $k=Np$, with a width that is of the
order of $\sqrt{N}$.  Over this range of $k$, the factor
$\left[1-(1\!-\!q)^k\right]^{N-k}$ in Eq.~\eqref{CN} is nearly constant.  We
therefore replace $k$ by its most probable value $Np$ in the above
expression.  After doing so,  this factor can be written as
\begin{align}
C(N)&\simeq\left[1-(1-q)^{Np}\right]^{N(1-p)} \nonumber \\
  &\simeq \exp\Big\{\!\!-N(1-p) \exp\big[N p \ln(1-q)\big]\Big\}\,. \nonumber
\end{align}

The probability that the network of $N$ nodes is complete, $\mathcal{C}(N)$,
is then given by
\begin{align}
\label{complete}
\mathcal{C}(N)&=\prod_{j\leq N-1} C(j)\nonumber\\
&\simeq \exp\Big\{\!\!-\!\int^N \!\!\!j(1\!-\!p) \exp\!\big[j\, p \ln(1\!-\!q)\big]\,dj\Big\}\,. 
\end{align}
Because the integral in the exponent converges as $N\to\infty$, the
completion probability is necessarily non-zero.

\begin{figure}
\includegraphics[width=0.4\textwidth]{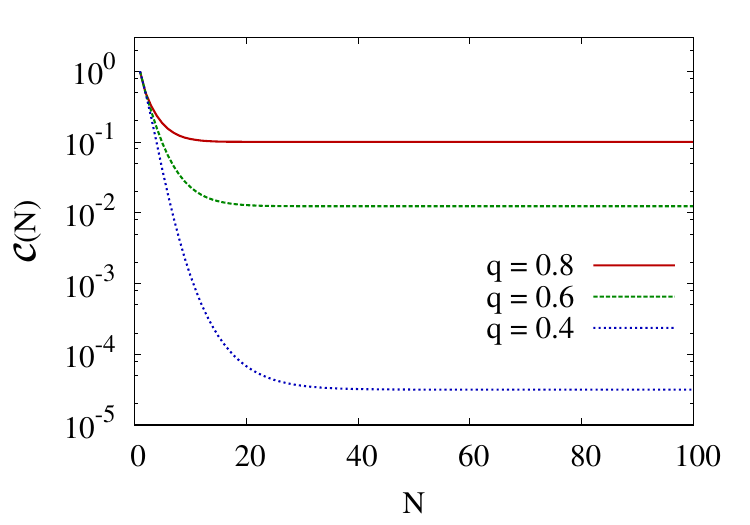}
\caption{Numerical evaluation of the probability for network completeness,
  $\mathcal{C}(N)$ from Eq.~\eqref{complete}, for fixed $p=\frac{1}{2}$ and representative values of
  $q$.  The saturation is obvious, $\mathcal{C}(\infty)>0$ for all $q>0$. The ultimate
  values of $\mathcal{C}(\infty)$ can be very small for small $q$.}
\label{fig:PiN}
\end{figure}

\begin{figure}
\includegraphics[width=0.4\textwidth]{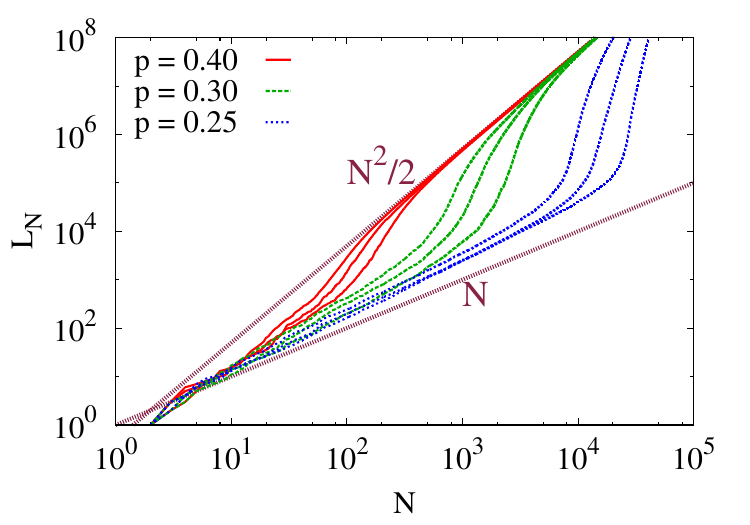}
\caption{The $N$ dependence of the number of links for second-neighbor
  copying with $q=p^2$.}
\label{L-vs-N}
\end{figure}

Numerical numerical evaluation of \eqref{complete} shows that the completion
probability $\mathcal{C}(N)$ indeed converges to a non-zero, albeit extremely
small, value as $N\to\infty$.  Figure~\ref{fig:PiN} shows this evaluation for
the case of $p=\frac{1}{2}$ and various $q$.  A more relevant criterion is
not defect-free completeness, but whether the number of links eventually
scales as $N^2/2$, as in the complete graph.  Simulations show that for
representative values of $p$ and $q$, the average number of links $L(N)$
initially grows linearly with $N$ but then crosses over to growing as $N^2/2$
(Fig.~\ref{L-vs-N}).  Thus second-order copying generically leads to networks
that are effectively complete---eventually each individual knows almost
everybody.  Moreover, Fig.~\ref{L-vs-N} illustrates the macroscopic
differences between individual network realizations.  Thus copying leads to
non-self-averaging in the dense regime---unpredictable outcomes when starting
from a fixed initial state.  This intriguing feature also arises in empirical
networks and related systems~\cite{SDW06,RW12,O14}, and intellectually
originates with the classic P\'olya urn model~\cite{M17,EP23,M09}.

\section{Outlook}
\label{concl}

We introduced and investigated the properties an exceedingly simple growing
network model that is based on the mechanism of node copying.  Each new node
that joins the network attaches to a randomly selected target node and also
to each of the neighbors of the target with an independent copying
probability $p$.  In spite of its deceptive simplicity, the structure of the
network that results from this growth mechanism is extremely rich.  One of
the fundamental outcomes of our copying model is that a transition from a
sparse to dense regime occurs as the copying probability $p$ increases beyond
$\frac{1}{2}$.  Dense networks are characterized by a mean degree that increases with
the number $N$ of nodes in the network, a feature that appears in a variety
of empirical networks~\cite{kleinberg}, as well as by large fluctuations
between individual network realizations.  For these reasons alone, it is
important to understand this densification process.  

The bulk of our analysis focused on the analytical description of various
global network quantities, such as the $N$ dependence of the average number
of links $L(N)$ and the number of $m$-cliques, $K_m(N)$.  We found that
$L(N)\sim N$ for $p<\frac{1}{2}$ and $L(N)\sim N^{2p}$ for $p>\frac{1}{2}$.
Analogously, for triangles, we found that $K_3(N)\sim N$ for $p<\frac{1}{2}$,
$K_3(N)\sim N^{2p}$ for $\frac{1}{2}<p<\tfrac{2}{3}$, and
$K_3(N)\sim N^{3p^2}$ for $p>\tfrac{2}{3}$.  For general $m$, there are $m-1$
transitions points where the $N$ dependence of the $m$-clique density
suddenly changes.  Given the richness of our predictions, it would be
worthwhile to reanalyze the densifying networks have have been observed
empirically~\cite{kleinberg} to test whether they can be accounted for within
the framework of the copying model.

The possibility to generate networks with controllable densities of specific
motifs might help in the design of controlled environments to explore how the
network topology affects the diffusion of an innovation or the spread of a
virus in a social system.  The copying model could also serve as benchmark to
test the veracity and the robustness of various types of algorithms for
characterizing networks, such as community detection~\cite{mod1,mod2}, by
generating more realistic structural properties than those of random
benchmarks~\cite{LFR08,PSHM13}.  Another basic unanswered question is: What
are the spectral properties of networks generated by the copying model?  This
question is particular intriguing in the dense regime where there are
large fluctuations between individual network realizations.

\newpage
\bigskip
{\bf Acknowledgments}. Financial support for this research was provided in
part by the grants from the ARC and the Belgian Network DYSCO, funded by the
IAP Programme (RL), DMR-1608211 and 1623243 from the National Science
Foundation (UB and SR), the John Templeton Foundation (SR), and Grant No.\
2012145 from the US-Israel Binational Science Foundation (UB).

\appendix
\section{Exact behavior of $L(N)$}
\label{app:LN}

To determine the exact solution of Eq.~\eqref{LN-exact}, we first solve the
homogeneous version of this equation and use this solution as an integrating
factor.  The homogeneous solution is
\begin{equation*}
\prod_{j=1}^{N-1}\Big(1+\frac{2p}{j}\Big)=\frac{\Gamma(2p+N)}{\Gamma(2p+1)\,\Gamma(N)}\,.
\end{equation*}
We thus we seek a solution to Eq.~\eqref{LN-exact} of the form
\begin{equation*}
L(N)=U(N)\,\frac{\Gamma(2p+N)}{\Gamma(2p+1)\,\Gamma(N)}\,.
\end{equation*}
This ansatz allows us to recast Eq.~\eqref{LN-exact} into the recurrence
\begin{equation}
\label{UN-rec}
U(N+1)=U(N)+\frac{\Gamma(2p+1)\,\Gamma(N+1)}{\Gamma(2p+N+1)}\,.
\end{equation}
Solving Eq.~\eqref{UN-rec} subject to the initial condition $U(1)=0$ (recall that $L_1=0$),
we find
\begin{equation}
\label{LN-exact-sol}
L(N)=\frac{\Gamma(2p+N)}{\Gamma(N)}\sum_{j=2}^{N}\frac{\Gamma(j)}{\Gamma(2p+j)}.
\end{equation}
To determine asymptotic properties, we will often use the well-known feature
of the gamma function
\begin{equation}
\label{x-asymp}
\frac{\Gamma(2p+x)}{\Gamma(x)}\to x^{2p} \qquad\quad x\gg 1.
\end{equation}
When $p<\frac{1}{2}$, the sum on the right-hand side of \eqref{LN-exact-sol}
diverges.  Thus we can use \eqref{x-asymp} to give
\begin{equation*}
L(N)\to N^{2p}\sum_{j\leq N}j^{-2p}\to \frac{N}{1-2p}\,,
\end{equation*}
leading to the result quoted in \eqref{LN-sol}.  For $p=\frac{1}{2}$ the exact
solution to \eqref{LN-exact-sol} is
\begin{equation}
\label{LN-sol-1/2}
L(N)=N(H_N-1)\,,
\end{equation}
where $H_N=\sum_{1\leq j\leq N} j^{-1}$ is the $N^{\rm th}$ harmonic number.
From the asymptotics of the harmonic numbers~\cite{knuth} we obtain
\begin{equation}
\label{LN-sol-asymp}
L(N)=N(\ln N+\gamma-1)+\frac{1}{2}-\frac{1}{12 N}+\frac{1}{120 N^3}+\ldots
\end{equation}
where $\gamma=0.57721566\ldots$ is the Euler-Masceroni constant.  Keeping
only the leading term in Eq.~\eqref{LN-sol-asymp} gives the result quoted in
\eqref{LN-sol}.  For $p>\frac{1}{2}$, the sum on the right-hand side of
Eq.~\eqref{LN-exact-sol} converges.  Hence
\begin{equation}
\label{Ap:def}
L(N)\to N^{2p}\sum_{j=2}^\infty\frac{\Gamma(j)}{\Gamma(2p+j)}\equiv A(p)\,N^{2p}
\end{equation}
with $A(p)$ given by Eq.~\eqref{A}.  The sum on the right-hand side of
Eq.~\eqref{Ap:def} is found by specializing the identity \cite{knuth}
\begin{equation}
\label{knuthid}
\sum_{k=0}^\infty\frac{\Gamma(a+k)}{\Gamma(c+k)}=\frac{\Gamma(a)}{(c-a-1)\,\Gamma(c-1)}
\end{equation}
to $a=2, ~c=2p+2$. 

\section{Exact Behavior of $T(N)$}
\label{app:TN}

To find the amplitude $C(p)$ quoted in \eqref{TN-sol}, we need to solve the
recurrence~\eqref{tevolution}.  Following the same approach as that used for
the number of links, we first solve the homogeneous version
of~\eqref{tevolution} and use this  the homogeneous solution as an
integrating factor
\begin{equation}
\label{tri}
T (N) = R(N)\, \frac{\Gamma(3p^2 + N )}{\Gamma(N)}
\end{equation}
We use this substitution together with the exact solution \eqref{LN-exact-sol} to recast \eqref{tevolution}
into recurrence
\begin{equation*}
R(N+1) = R(N) + 2p \frac{\Gamma(2p + N)}{\Gamma(3 p^2 +1 +N)} \sum_{j=2}^N \frac{\Gamma(j)}{\Gamma(2p +j)}
\end{equation*}
which is solved to give
\begin{equation}
\label{twosums}
R(N) =  2p \sum_{j=2}^{N-1} \frac{\Gamma(j)}{\Gamma(2p +j)}  \sum_{n=j}^{N-1} \frac{\Gamma(2p + n)}{\Gamma(3 p^2 + 1 + n)}\,.
\end{equation}

When $p>\tfrac{2}{3}$, both sums in \eqref{twosums} are convergent.  Hence \eqref{tri}
asymptotically becomes $T(N) = R(\infty) N^{3 p^2}$, where we additionally
used \eqref{x-asymp}.  Thus the amplitude $C(p)$ in Eq.~\eqref{TN-sol} is
equal to $R(\infty)$; that is,
\begin{equation}
\label{summy}
C(p)=2p\sum_{j=2}^{\infty} \frac{\Gamma(j)}{\Gamma(2p +j)}
\sum_{n=j}^{\infty} \frac{\Gamma(2p+n)}{\Gamma(3p^2+1+n)}\,.
\end{equation}
Using the identity \eqref{knuthid} twice, we compute the sums in
\eqref{summy} and arrive at the result for $C(p)$ quoted in
Eq.~\eqref{C}.

\end{document}